\documentclass[pdflatex,sn-mathphys-num]{sn-jnl}

\usepackage{silence}
\ActivateWarningFilters[hyperreflevel]

\setcitestyle{numbers,square}   

\usepackage{etoolbox}
\apptocmd{\sloppy}{\hbadness 10000\relax}{}{}

\usepackage{lmodern}

\usepackage[group-separator={,},group-minimum-digits={3}]{siunitx}
\usepackage{afterpage}
\usepackage{dsfont}
\usepackage[utf8]{inputenc} 
\usepackage[T1]{fontenc}    
\usepackage{booktabs}       
\usepackage{url}  

\usepackage{breakurl}
\usepackage[breaklinks]{hyperref}
\usepackage{amsfonts}       
\usepackage{nicefrac}       
\usepackage{microtype}      
\usepackage{lipsum}
\usepackage{fancyhdr}       
\usepackage{graphicx}       
\graphicspath{{media/}}     
\usepackage{amsmath}
\usepackage{mathtools}
\usepackage[toc]{appendix}
\usepackage{minitoc}
\usepackage{caption}
\usepackage{subcaption}
\usepackage{textcomp}  
\usepackage{gensymb}
\usepackage{booktabs}
\usepackage{multirow}
\usepackage{float}
\usepackage{placeins}
\usepackage{tabularx}
\usepackage{makecell}
\usepackage{rotating}
\usepackage{setspace}
\usepackage{parskip}
\usepackage{newfloat}

\usepackage{xcolor} 
\usepackage{xr}
\usepackage[capitalise,noabbrev]{cleveref}

\DeclareFloatingEnvironment[
    fileext=efd,
    placement={!ht},
    name=Extended Data Figure
]{edfigure}

\DeclareFloatingEnvironment[
    fileext=efd,
    placement={!ht},
    name=Extended Data Table
]{edtable}

\DeclareCaptionLabelFormat{sif}{Figure S#2}
\DeclareFloatingEnvironment[
    fileext=efd,
    placement={!ht},
    name=Figure S
]{sifigure}

\DeclareCaptionLabelFormat{sit}{Table S#2}
\DeclareFloatingEnvironment[
    fileext=efd,
    placement={!ht},
    name=Table S
]{sitable}

\crefname{figure}{Fig.}{Figs.}
\crefname{edfigure}{Extended Data Fig.}{Extended Data Figs.}
\crefname{edtable}{Extended Data Table}{Extended Data Tables.}
\crefname{sifigure}{Fig. S}{Figs. S}
\creflabelformat{sifigure}{#2#1#3}
\crefname{sitable}{Tab. S}{Tabs. S}
\creflabelformat{sitable}{#2#1#3}

\crefname{section}{}{}

\crefname{appendix}{Supplementary}{Supplementaries}
\Crefname{appendix}{Supplementary}{Supplementaries}

\AtBeginEnvironment{appendices}{\crefalias{section}{appendix}}
\AtBeginEnvironment{appendices}{\crefalias{subsection}{appendix}}
\AtBeginEnvironment{appendices}{\crefalias{subsubsection}{appendix}}

\usepackage{enumitem}
\setlist[itemize]{leftmargin=1cm}

\renewcommand{\paragraph}[1]{\textbf{#1}}

\makeatletter
\let\@LN\relax
\makeatother

\usepackage{geometry}  
\begin{document}
\doparttoc
\faketableofcontents

\title[\textbf{Data-driven global ocean model resolving ocean-atmosphere coupling dynamics}]{\textbf{Data-driven global ocean model resolving ocean-atmosphere coupling dynamics}}

\author*[1]{\fnm{Jeong-Hwan} \sur{Kim}}\email{jeonghwan@kist.re.kr}
\author*[1]{\fnm{Daehyun} \sur{Kang}}\email{dkang@kist.re.kr}
\author[2]{\fnm{Young-Min} \sur{Yang}}
\author[3]{\fnm{Jae-Heung} \sur{Park}}
\author[4]{\fnm{Yoo-Geun} \sur{Ham}}

\affil[1]{\orgdiv{Center for Climate and Carbon Cycle Research, Korea Institute of Science and Technology, Seoul, Republic of Korea}}
\affil[2]{\orgdiv{Department of Environment and Energy, Jeonbuk National University, Jeonju, Republic of Korea}}
\affil[3]{\orgdiv{School of Earth and Environmental Sciences, Seoul National University, Seoul, Republic of Korea}}
\affil[4]{\orgdiv{Department of Environmental Management, Seoul National University, Seoul, Republic of Korea}}

\abstract{
Artificial intelligence has advanced global weather forecasting, outperforming traditional numerical models in both accuracy and computational efficiency. Nevertheless, extending predictions beyond subseasonal timescales requires the development of deep learning (DL)-based ocean-atmosphere coupled models that can realistically simulate complex oceanic responses to atmospheric forcing. This study presents KIST-Ocean, a DL-based global three-dimensional ocean general circulation model using a U-shaped visual attention adversarial network architecture. KIST-Ocean integrates partial convolution, adversarial training, and transfer learning to address coastal complexity and predictive distribution drift in auto-regressive models. Comprehensive evaluations confirmed the model’s robust ocean predictive skill and efficiency. Moreover, it accurately captures realistic ocean responses, such as Kelvin and Rossby wave propagation in the tropical Pacific, and vertical motions induced by cyclonic and anticyclonic wind stress, demonstrating its ability to represent key ocean-atmosphere coupling mechanisms underlying climate phenomena, including the El Niño-Southern Oscillation. These findings reinforce confidence in DL-based global weather and climate models and their ability to represent Earth’s complex interactions. This study lays a foundation for extending DL-based approaches to broader Earth system modeling, offering potential for enhancing climate prediction capabilities.
}


\begingroup
\thispagestyle{empty}
\setlength{\parskip}{0pt}  
\maketitle
\endgroup

\section{Introduction}
\label{sec:intro}

With the advent of the artificial intelligence (AI) era, the application of deep learning (DL) in global weather prediction has yielded remarkable achievements\cite{Bi2023, Lam2023, Chen2024, Chen2023a, Chen2023b, Ling2024, Pathak2022}. These DL-based weather prediction models have surpassed traditional state-of-the-art (SOTA) numerical weather prediction models in terms of both forecast skill and computational efficiency\cite{Bi2023, Lam2023, Chen2023a, Chen2023b, Rasp2024}. However, several challenging issues remain, such as extending lead times beyond seasonal scales and improving the prediction of tropical cyclone intensity and extreme climate events (e.g., heat waves, and extreme precipitation)\cite{Schultz2021, Bouallgue2024, Liu2024, DeMaria2024}. Among these challenges, extending the prediction horizon is particularly significant, as it provides societies with more time to prepare for natural disasters and delivers substantial socioeconomic value. Beyond global weather prediction, it is now imperative to further investigate the application of DL in global climate prediction. To advance this objective, increased focus should be directed toward ocean modeling.

The ocean, which covers the majority of the Earth’s surface, has a substantially higher heat capacity than the atmosphere, resulting in slower temperature changes. This ocean memory effect is a vital source of long-term predictability\cite{Deser2003, Srivastava2017}. The persistence of these oceanic states has been effectively utilized in the development of DL-based climate prediction models, enabling significant advancements in the multi-year forecasting of phenomena such as the El Niño-Southern Oscillation (ENSO)\cite{Ham2019, Zhou2023} and the Indian Ocean Dipole\cite{Ling2022}. Moreover, the interaction between the ocean and atmosphere, where energy, momentum, and carbon fluxes are exchanged, plays a pivotal role in the Earth’s system\cite{Large2009}. These complex processes significantly influence global weather and climate, underpinning large-scale climate variability mechanisms\cite{Newman2009}.

Recognizing this importance, traditional dynamical climate prediction systems have adopted a strategy of simulating atmosphere-ocean coupled models. Owing to the distinct characteristics of the ocean and atmosphere, these models integrate separate ocean general circulation models (OGCMs) and atmospheric general circulation models using a coupler that exchanges fluxes at their boundaries\cite{Merryfield2013, Saha2014, Molod2020}. This approach reflects the fundamental complexity of the coupled ocean-atmosphere system, while enabling enhanced long-term prediction skills.

Similarly, for DL models to successfully expand their application from weather to climate prediction, the development of an ocean-atmosphere coupled model is essential, with the first step being ocean modeling. Recently, several DL-based global three-dimensional (3D) ocean models have been developed\cite{Xiong2023, Wang2024a, Guo2024a, Wang2024b}. Some of these models have demonstrated improved predictive skills compared to those of operational ocean forecasting systems\cite{Wang2024a} or dynamical ocean-atmosphere coupled models\cite{Guo2024a}. However, they are often constrained by non-autoregressive forecasting approaches—either constructing separate models for each lead time, which results in discontinuous predictions\cite{Wang2024a}, or using a monthly temporal interval, the temporal resolution of which is too coarse to capture an ocean model’s response to surface boundary forcing\cite{Guo2024a}.

A recent attempt to implement a DL-based ocean-atmosphere coupled model reproduced realistic Kelvin and Rossby waves in the tropical Pacific that were comparable to the ground truth\cite{Wang2024b}. However, because this model represents the ocean using only potential temperature, sea surface temperature (SST), and sea surface height, it lacks prognostic currents and salinity, limiting its ability to resolve subsurface circulation and heat content redistribution. Moreover, as no dedicated atmospheric forcing experiments have been conducted, it remains unclear whether the simulated Kelvin and Rossby waves were generated by the wind stress or arose from the initial ocean state.

These limitations make it challenging to assess whether DL-based ocean models can provide physically valid responses to boundary conditions, raising the question of whether they can effectively simulate ocean-atmosphere interactions. In this study, we address this uncertainty by developing a DL-based OGCM and evaluating its physical consistency more extensively under various atmospheric forcing experiments. To the best of our knowledge, this is the first study to verify oceanic responses driven by the atmosphere, such as tropical waves and Ekman divergence, against predictions from traditional theories.

\section{Global Ocean Modeling with KIST-Ocean}

\begin{figure}[h!]
	\centering
	\includegraphics[width=0.9\textwidth]{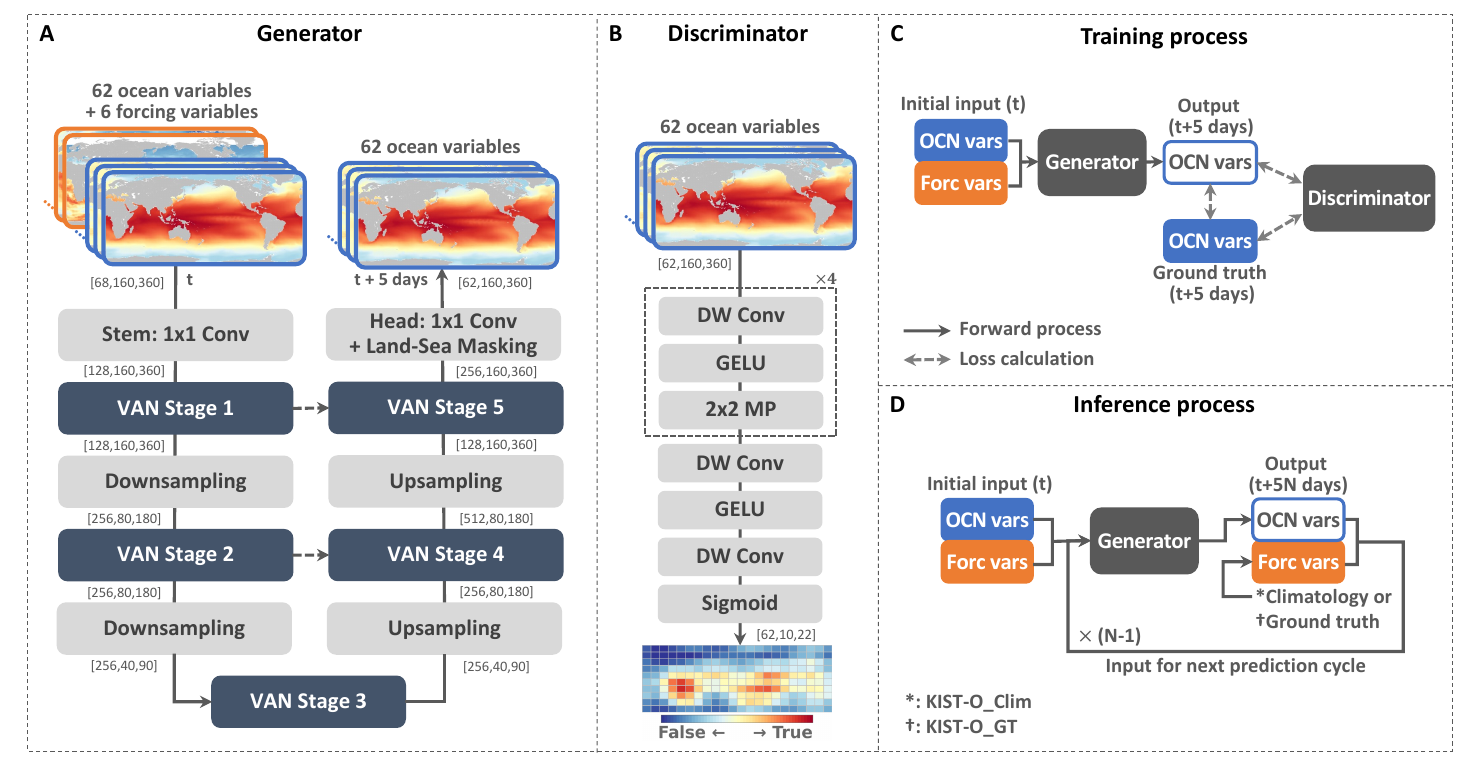}
	\caption{\textbf{Overview of the Korea Institute of Science and Technology’s Ocean model (KIST-Ocean), including its training and inference processes. A}, Generator architecture. For details on the visual attention network (VAN) stages, down-sampling blocks, and up-sampling blocks, see Extended Data Fig. 1. \textbf{B}, Discriminator architecture. DW Conv and MP denote depth-wise convolution and max pooling, respectively, while GELU indicates the Gaussian Error Linear Unit activation function. \textbf{C}, Training process. The model was trained using a generative adversarial network (GAN) framework, in which the generator and discriminator learn in competition. The generator receives ocean and surface boundary forcing variables (denoted as OCN vars and Froc vars, respectively) and predicts ocean variables five days ahead. \textbf{D}, Inference process. The figure illustrates how KIST-Ocean generates future ocean states after training is complete. Maps were generated using the Basemap Toolkit (v1.2.0).}
	\label{fig:fig1}
\end{figure}

In this study, we propose a U-shaped visual attention adversarial network-based data-driven model, the Korea Institute of Science and Technology’s Ocean model (KIST-Ocean), designed to simulate the global 3D ocean. We designed KIST-Ocean as a component of a coupled ocean-atmosphere model that produces only oceanic variables, similar to dynamical OGCMs (\cref{fig:fig1}A). KIST-Ocean integrates 62 oceanic variables along with 6 surface boundary conditions as inputs to generate predictions for oceanic variables (\cref{edtab:ed_tab1}; \cref{sec:text_s1}). The prediction time interval was five days; therefore, both the input and output variables were averaged over five days.

KIST-Ocean builds on three key innovations. The first involves the integration of a visual attention network (VAN)\cite{Guo2023, Li2023}. VAN decomposes convolution to achieve larger receptive fields with fewer parameters. Combining the U-shaped architecture with VAN further enhances efficiency. The U-shaped design captures multi-scale features and global context through dimension reduction and restoration. Meanwhile, skip connections integrate local and global features, ensuring robustness with limited data\cite{Ronneberger2015}. This efficient design enables KIST-Ocean to simulate the global 3D ocean with 6.6 million parameters, requiring approximately 33.3 h for pretraining and 2.4 h for fine-tuning on a single NVIDIA A100 GPU.

The second innovation is the application of a partial convolution. The gridded oceanic datasets include land-ocean interface, complicating DL applications. Convolutional neural networks share kernels horizontally\cite{Goodfellow2016}, potentially underestimating coastal variability. By incorporating partial convolution—a method that excludes masked values—we prevented distortions caused by land grid cells. This approach enabled the model to more accurately capture the complex variability in coastal\cite{Liu2018, Ham2024}.

The third innovation involves the application of adversarial training, which addresses distributional drift—a common issue in auto-regressive DL models that reuse outputs as inputs in subsequent inference steps. We employed a conditional generative adversarial network (GAN) framework to better align predictions with ground truth\cite{Li2023, Ham2024}, utilizing PatchGAN\cite{Isola2017} discriminator that independently evaluates multiple image patches (\cref{fig:fig1}B). These methodologies are described in detail in the Methods section.

Consequently, KIST-Ocean was trained through competitive training of the generator and discriminator (\cref{fig:fig1}C). To ensure sufficient training data and stable model training, we employed a transfer learning. First, the model was pre-trained using long-term simulation data from the Community Earth System Model version 2 (CESM2) Large Ensemble dataset. The model was then fine-tuned with decades of reanalysis datasets (1982--2013) (\cref{edtab:ed_tab2}).

During inference, KIST-Ocean operates auto-regressively (\cref{fig:fig1}D), generating forecasts up to 200 d into the future, which requires approximately 6--7 s on a single NVIDIA A100 GPU. Because KIST-Ocean does not output surface boundary conditions, these must be prescribed either from ground truth (KIST-O\_GT) or climatology (KIST-O\_Clim), representing the upper and lower bounds of model’s forecast capability, respectively. To evaluate the performance of KIST-Ocean, we conducted hindcast experiments spanning 2014--2023, comparing its results against persistence forecasts and dynamical seasonal predictions from the North American Multi-Model Ensemble models (NMMEs).

\section{Evaluating KIST-Ocean’s Performance}

\begin{figure}[h!]
	\centering
	\includegraphics[width=0.9\textwidth]{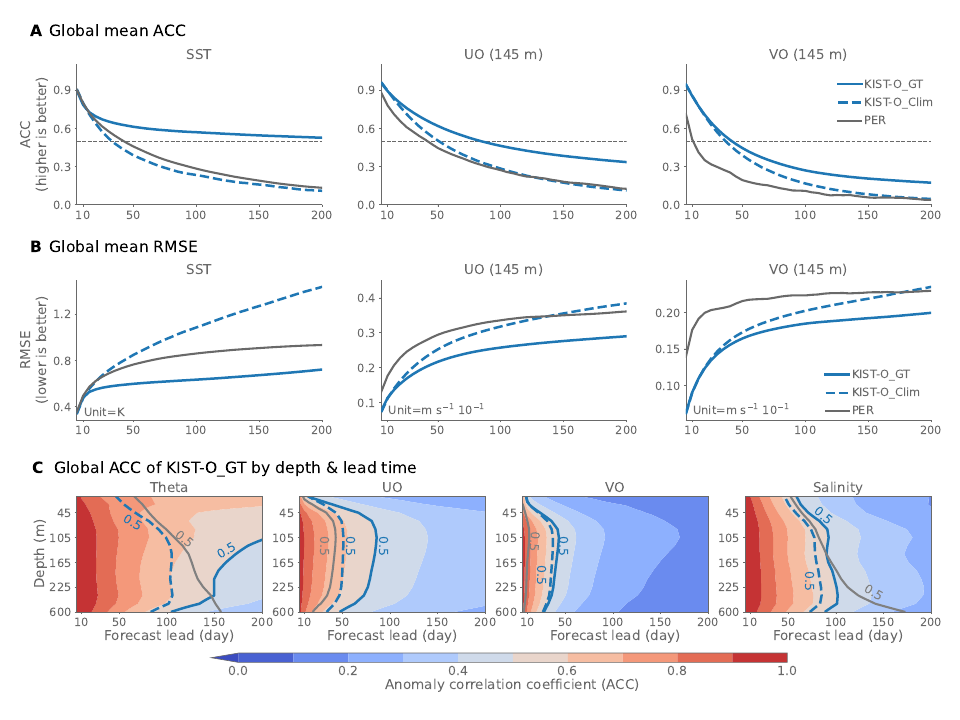}
	\caption{\textbf{Comparative evaluation of KIST-Ocean’s global three-dimensional (3D) ocean prediction skill against that of a persistence forecast. A}, Globally averaged anomaly correlation coefficient (ACC) skill for sea surface temperature (SST), zonal current (UO) at a depth of 145 m, and meridional current (VO) at a depth of 145 m (left to right), computed over 2014--2023. The solid blue and dashed blue lines represent KIST-O\_GT (with ground truth prescribed as the surface boundary forcing) and KIST-O\_Clim (with climatology prescribed), respectively. The solid gray line denotes the persistence forecast, and thin gray dashed lines indicate the ACC = 0.5 threshold. \textbf{B}, Same as A, but for globally averaged root mean square error (RMSE). \textbf{C}, Globally averaged ACC of KIST-O\_GT for all 3D ocean variables, presented as a function of depth and lead time. From left to right: potential temperature (Theta), UO, VO, and salinity. Shading represents the ACC of KIST-O\_GT, while contours indicate the ACC = 0.5 threshold for KIST-O\_GT (solid blue), KIST-O\_Clim (dashed blue), and the persistence forecast (solid gray).}
	\label{fig:fig2}
\end{figure}

When evaluated over 200 d global ocean forecasts from 2014 to 2023, KIST-Ocean demonstrated robust ocean prediction skills relative to that of a persistence forecast. \cref{fig:fig2} shows that, by prescribing ground truth surface boundary forcing (i.e., KIST-O\_GT), the model outperformed persistence forecasts for key variables, including SST, zonal current (UO) at 145 m, and meridional current (VO) at 145 m, in terms of both the anomaly correlation coefficient (ACC) and root mean square error (RMSE). Specifically, KIST-O\_GT achieved significant ACC for potential temperature, UO, VO, and salinity. The maximum lead times with significant skill were up to 200, 85, 40, and 100 d, respectively. Overall, among the 2,480 prediction targets (62 variables $\times$ 40 lead times), KIST-Ocean surpassed the persistence forecast for 81.3\% (ACC) and 86.5\% (RMSE) of targets (\cref{edfig:ed_fig2}A and \cref{edfig:ed_fig3}A).

The SST forecasts from KIST-O\_Clim, which prescribe climatological boundary forcing without observational inputs, exhibited a rapid increase in RMSE compared to KIST-O\_GT, reflecting the significant influence of surface boundary forcing. The substantial skill gap between KIST-O\_GT and KIST-O\_Clim highlights the model’s effective utilization of surface boundary information. KIST-O\_GT retained the longest-lived skill for near-surface potential temperature, indicating that KIST-Ocean effectively learned the distinct relationships between boundary forcing and vertical ocean structure. Conversely, KIST-O\_Clim maintained comparatively longer skill in subsurface currents, outperforming the persistence forecast for lead times under 120 d, suggesting subsurface initial conditions provide additional predictive capacity. Moreover, the model successfully reproduced the hierarchical relationships among ocean variables with depth, even without explicit depth embeddings.

\begin{figure}[h!]
	\centering
	\includegraphics[width=0.9\textwidth]{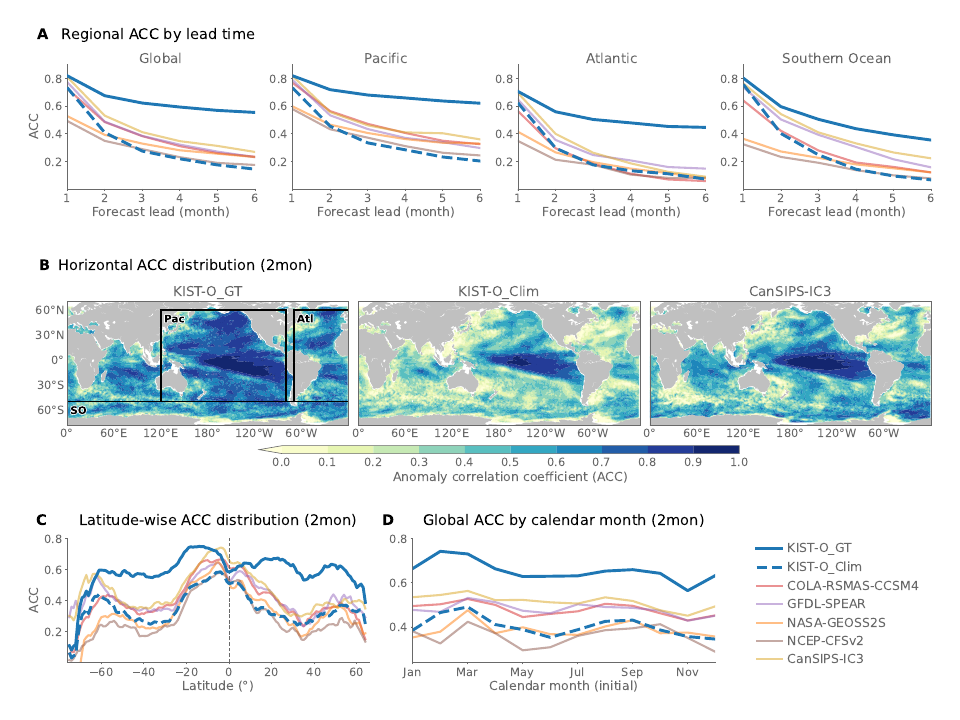}
	\caption{\textbf{Comparative evaluation of ACC for monthly SST between KIST-Ocean and the North American Multi-Model Ensemble (NMME) models. A}, Regional mean ACC as a function of lead time (months) for the globe, Pacific (120°E--80°W, 50°S--60°N), Atlantic (70°W--0°, 50°S--60°N), and Southern Ocean (0°--360°, 50°--79°S), shown from left to right. The colored lines represent KIST-O\_GT (solid blue), KIST-O\_Clim (dashed blue), COLA-RSMAS-CCSM4 (red), GFDL-SPEAR (purple), NASA-GEOSS2S (orange), NCEP-CFSv2 (brown), and CanSIPS-IC3 (yellow). \textbf{B}, Horizontal distributions of two-month-lead ACC for KIST-O\_GT, KIST-O\_Clim, and CanSIPS-IC3, computed over 2015--2022 (left to right). \textbf{C}, Latitudinal distribution of ACC at a two-month lead. The x-axis represents latitude (°), while the y-axis represents ACC. D, ACC at a two-month lead time by initial forecast month. The x-axis represents the initial month (calendar month), while the y-axis represents ACC. Maps were generated using the Basemap Toolkit (v1.2.0).}
	\label{fig:fig3}
\end{figure}

To further evaluate the predictive skill of KIST-Ocean, we compared its monthly SST forecasts against those from the NMMEs. When provided with ground truth forcing (KIST-O\_GT), KIST-Ocean surpassed the NMMEs in predicting SST for the global, Pacific, Atlantic, and Southern Ocean up to a six-month lead time (\cref{fig:fig3}A), reaffirming its capability to effectively utilize surface boundary conditions. Even under climatological forcing (KIST-O\_Clim), the model’s one-month ACC matched that of the highest-performing NMMEs, remaining within their skill range at two-month lead times, and notably extending up to six months in the Atlantic Ocean. Hence, KIST-Ocean maintained stable SST predictive skill for one- to two-month lead times, regardless of forcing quality (\cref{fig:fig3}B).

KIST-Ocean achieved robust SST forecasts across most latitudes, although the skill gap between KIST-O\_GT and KIST-O\_Clim was relatively pronounced within approximately 20--50$^{\circ}$ latitude bands (\cref{fig:fig3}C). This gap diminished near the equator and Antarctica, due to the predominance of subsurface processes associated with ENSO and the Antarctic Circumpolar Current, respectively. While the predictive skill of KIST-Ocean showed minimal seasonal dependence, forecasts initialized in late boreal winter performed better than those initiated in summer or autumn—a pattern also observed in the NMMEs (\cref{fig:fig3}D). This likely results from a deeper ocean mixed layer and enhanced thermal memory in the tropical Pacific during late winter\cite{Kirtman2014, Barnston2019}. Collectively, these evaluations demonstrate KIST-Ocean’s capability to deliver reliable multi-month forecasts of the global ocean.

\section{Evaluating Physical Coupling Processes}

\begin{figure}[h!]
	\centering
	\includegraphics[width=0.9\textwidth]{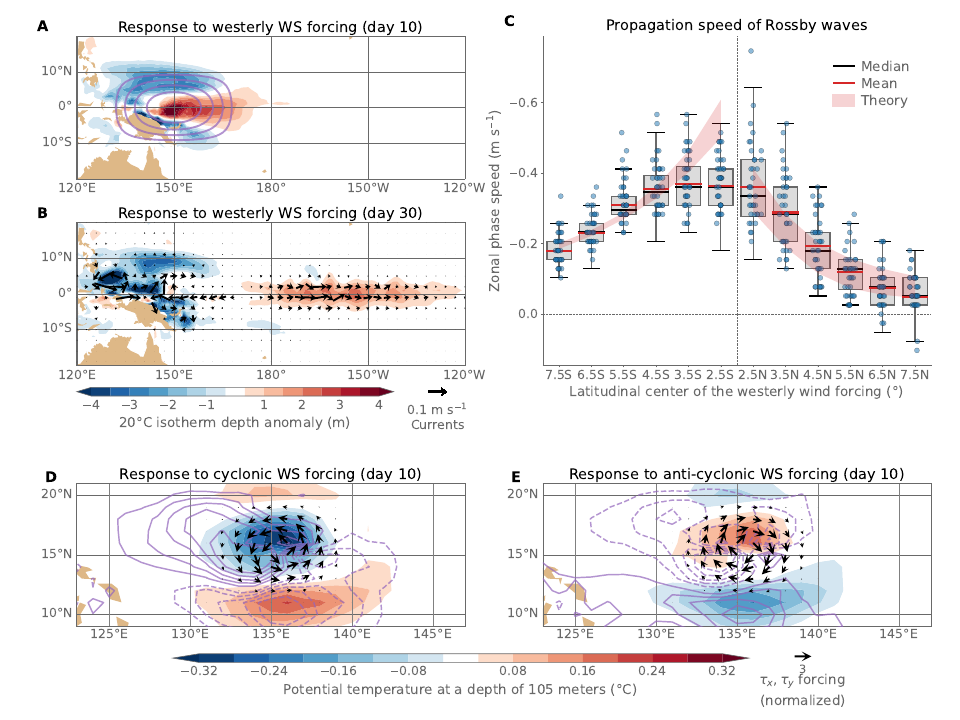}
	\caption{\textbf{KIST-Ocean’s dynamical response to idealized wind stress forcing. A} and \textbf{B}, Tropical Pacific response to a bell-shaped westerly wind burst (WWB) centered at 130$^{\circ}$--170$^{\circ}$E, 10$^{\circ}$S--10$^{\circ}$N. Shading denotes the 20$^{\circ}$C isotherm depth anomaly (m), purple contours indicate the normalized zonal wind stress forcing (unitless), and black arrows show depth-averaged (5--65 m) horizontal currents (m s$^{-1}$). Snapshots are shown at lead days 10 (A) and 30 (B). The initial ocean state corresponds to December 12--16, 2013. \textbf{C}, Propagation speed of Rossby waves generated by WWBs at different latitudes. Boxes represent the interquartile range; whiskers extend to 1.5 times this range; black and red lines denote the median and mean, respectively. Light-red shading indicates the theoretical long-wave phase speed derived from the first baroclinic Rossby wave approximation. The x-axis denotes the central latitude of the WWB ($^{\circ}$), and the y-axis indicates the Rossby wave zonal phase speed (m s$^{-1}$). \textbf{D} and \textbf{E}, Western North Pacific response (131$^{\circ}$--140$^{\circ}$E, 11$^{\circ}$--20$^{\circ}$N) at lead day 10 to cyclonic (D) and anticyclonic (E) rotational wind stress forcing. Shading represents the potential temperature anomaly at 105 m ($^{\circ}$C). Black arrows indicate the normalized surface wind stress forcing (unitless), and purple contours denote the vertical current speed inferred from mass continuity, with an interval of 0.5 ($10^{-6}$ m s$^{-1}$). Experimental details are provided in “Idealized Wind Stress Forced Experiments” of the Methods. Maps were generated using the Basemap Toolkit (v1.2.0).}
	\label{fig:fig4}
\end{figure}

Assessing whether KIST-Ocean has realistically learned the physical relationships between surface boundary conditions and ocean variables is essential. Addressing this issue allows us to evaluate the consistency of KIST-Ocean’s physical processes and provides insight into the feasibility of implementing a DL-based ocean--atmosphere coupled model. Accordingly, we conducted nudging experiments prescribing surface wind forcing to KIST-Ocean to evaluate its ability to reproduce two representative physical ocean responses to atmospheric forcing: oceanic wave responses and Ekman transport (detailed in Methods).

In the tropical Pacific, downwelling oceanic Kelvin waves triggered by westerly wind bursts (WWBs) promote El Niño development by suppressing the upwelling of subsurface cold water in the eastern Pacific\cite{Timmermann2018}. To assess KIST-Ocean’s capability to simulate these wave dynamics, we prescribed westerly wind stress over the western Pacific and analyzed anomalies in the depth of the 20$^{\circ}$C isotherm as an indicator of thermocline variations\cite{Herbert2018}. KIST-Ocean realistically reproduced the eastward-propagating downwelling Kelvin waves and the slower westward-propagating upwelling Rossby waves (\cref{fig:fig4}A, B). The model also successfully captured the reflection of upwelling Rossby waves off the Maritime Continent as equatorial upwelling Kelvin waves, accurately representing the associated subsurface currents and the opposite-phase responses to easterly wind forcing (\cref{edfig:ed_fig4}. 4). Consequently, KIST-Ocean demonstrates model fidelity in realistically simulating response to WWBs.

Additional hindcasts initialized on December 14, 2013, and May 3, 2015, further support this conclusion (\cref{sec:text_s2} and \cref{sifig:si_fig1,sifig:si_fig2,sifig:si_fig3,sifig:si_fig4}) In 2013 case, KIST-O\_GT successfully reproduced the observed WWB-induced downwelling Kelvin waves and the subsequent SST warming, whereas KIST-O\_Clim did not. In 2025 case, realistic WWBs enabled KIST-O\_GT to accurately capture the Niño3.4 evolution into the strong El Niño, whereas KIST-O\_Clim rapidly discharged heat and evolved toward La Niña conditions. These contrasting outcomes highlight the model’s proficiency in simulating wave dynamics, recharge--discharge processes and Bjerknes feedback essential to ENSO evolution.

To further assess KIST-Ocean’s representation of key physical properties associated with latitude and thermocline depth, we prescribed westerly wind forcing at varying latitudes in the Eastern Pacific and analyzed changes in Rossby wave propagation speed (\cref{edtab:ed_tab3} and \cref{edfig:ed_fig6}). The propagation speed of Rossby waves is expected to decrease with stronger Coriolis forces at higher latitudes, whereas it increases with deeper thermoclines.

Remarkably, KIST-Ocean produced physically consistent results, showing slower Rossby wave propagation speeds at higher latitudes and faster propagation in the Southern Hemisphere compared to the Northern Hemisphere (\cref{fig:fig4}C). This behavior aligns with the relationship between Rossby wave speed and thermocline depth, which is greater in the southern tropical Pacific (\cref{edfig:ed_fig5}). In particular, the simulated latitudinal variations in Rossby wave speeds closely matched theoretical phase speeds via long-wave approximation (detailed in Methods)\cite{Killworth1997}. These results indicate that KIST-Ocean effectively learned governing wave dynamics.

Strong cyclonic wind stress forcing, such as that associated with tropical cyclones, induces Ekman transport, pushing surface water outward and drawing up cooler, deeper water. Consequently, cyclonic circulation centers cool, whereas anticyclonic circulation centers warm, influencing local cloud formation, precipitation, and tropical cyclone development\cite{Ma2020}. Evaluating whether a model can reproduce this mechanism provides insight into how effectively it captures the physical processes governing vertical oceanic motion and the associated temperature changes. Accordingly, we performed experiments prescribing cyclonic and anticyclonic wind stress forcing in KIST-Ocean.

Although KIST-Ocean does not explicitly formulate vertical velocity, vertical motion can be inferred from horizontal currents via the continuity equation\cite{Wang2001} (Text S3). KIST-Ocean realistically generated pronounced upwelling and surface cooling at the center of cyclonic wind forcing, and corresponding downwelling and warming under anticyclonic conditions (\cref{fig:fig4}D, E). Warm (cold) temperature anomalies developed to the north and south of the cyclonic (anticyclonic) forcing center as the divergent (accumulated) water moved away from the upwelling (downwelling) zone, inducing downwelling (upwelling) in these regions.

The ability of KIST-Ocean to reproduce the upwelling-downwelling forced by cyclonic and anticyclonic wind stress underpins its skill in simulating larger-scale coupled phenomena, such as ENSO. Accurate representation of Ekman divergence and convergence, along with vertical heat redistribution, is essential for capturing the initiation and evolution of equatorial waves, the recharge-discharge of warm water volume, and the activation of the Bjerknes feedback—all core elements of ENSO\cite{Timmermann2018}. Thus, the realistic cold wake response to cyclonic forcing and warming under anticyclonic forcing provide evidence that KIST-Ocean has learned the coupled ocean-atmosphere processes critical for ENSO cycle.

\section{Sensitivity Experiments}

To enhance KIST-Ocean’s modeling capabilities, we implemented partial convolution, adversarial training, and transfer learning, systematically evaluating each component’s individual contribution. We also compared KIST-Ocean’s performance with a FourCastNet-based model to assess the impact of different DL algorithms on global 3D ocean forecasting\cite{Kurth2023}. \cref{edfig:ed_fig6}A illustrates substantial improvements from partial convolution and adversarial training; omitting either led to a rapid ACC decline below 0.5 within 50 d for key variables. Transfer learning had a more modest overall impact but was increasingly beneficial at deeper layers, particularly for potential temperature and salinity (\cref{edfig:ed_fig7}; \cref{sifig:si_fig7}), likely due to lower inter-sample independence and heightened sensitivity to sample size.

To assess the realism of the predicted distributions, we examined the relative probability density functions of each model’s 60-d forecasts relative to the ground truth (\cref{edfig:ed_fig6}B). Models lacking partial convolution or adversarial training exhibited clear distribution drift, with excessive extreme values and underrepresentation of near-median values. The horizontal distributions of subsurface ocean current speed (\cref{edfig:ed_fig6}C) illustrate this drift. While the original KIST-Ocean closely matched ground truth, the version without partial convolution produced artificially strong currents across most tropical-subtropical oceans, and the version without adversarial training exhibited pronounced blurring, unrealistically weakening equatorial and western boundary currents.

Although the FourCastNet-based model included 25 million parameters, nearly 4 times that of KIST-Ocean (6.6 million), it performed notably worse in predictive skill and distribution alignment. This was especially evident for subsurface zonal currents, where the probabilities of near-median values decreased and extreme values increased, suggesting that specialized DL designs are required for effective global ocean modeling.

These experiments underscore the importance of partial convolution for modeling variables with extensive missing points, such as ocean or land surface variables. Additionally, adversarial training effectively mitigates distribution drift in auto-regressive models, enhancing overall performance. Meanwhile, transfer learning had a modest impact, suggesting that KIST-Ocean can achieve robust ocean modeling skill with a relatively limited sample size. This finding highlights the efficiency of the visual attention adversarial network design for global ocean modeling, particularly predicting of data-limited phenomena.

\section{Discussion}\label{sec:discussion}

This study presents KIST-Ocean, an efficient DL-based OGCM with 6.6 million parameters built on a U-shaped visual attention adversarial network. Although currently limited to ocean simulations, KIST-Ocean demonstrates robust global 3D ocean forecasts under realistic surface boundary conditions and delivers SST predictions for up to two months comparable to dynamic coupled models, regardless of boundary condition quality. In particular, KIST-Ocean significantly outperformed persistence forecasts in capturing challenging subsurface flows without realistic surface forcing. However, because the assessments were made under prescribed boundary conditions, additional work with a fully coupled model will be required to determine whether the present DL-based ocean model can translate into improved long-term prediction skill.

Considering the extensive interactions among Earth system components (e.g., atmosphere, ocean, sea ice, and land), it is crucial to assess whether DL-based weather and climate models can replicate these interactions with physical consistency. This is pivotal for envisioning a DL-based Earth system model capable of long-term climate prediction and supporting policy-relevant responses to climate change. Accordingly, we performed wind stress forcing experiments to examine whether KIST-Ocean realistically reproduces oceanic responses to atmospheric perturbations.

KIST-Ocean successfully simulated Kelvin and Rossby waves in the tropical Pacific, accurately capturing their eastward and westward propagations, respectively. Beyond these fundamental features, the model also replicated more nuanced dynamics, such as the varying propagation speeds of Rossby waves and the vertical oceanic motions induced by rotational wind stress forcing. These results confirm that KIST-Ocean learned the complex physical linkages between atmospheric forcing and ocean responses, demonstrating that DL-based approaches can realistically capture the physical interactions among components of the Earth system. The findings further suggested that developing a DL-based atmosphere-ocean coupled model may be both feasible and promising. Although the perturbed experiments yielded qualitatively reliable results, further quantitative assessments are warranted in future studies.

Finally, the U-shaped visual attention adversarial network architecture proved highly effective despite limited sample sizes and parameters. Partial convolution and adversarial training effectively incorporated the heterogeneous land-sea mask and varying ocean depths, offering valuable guidance for future DL-based Earth system modeling. Moreover, the sensitivity experiments provide important insights for future efforts in developing DL-based coupled models, potentially enhancing our ability to model intricate ocean-atmosphere interactions.

Our comprehensive study of the ocean-atmosphere coupling capability of a DL-based model demonstrates that it exhibits realistic dynamical properties and reliable predictability for ocean-atmosphere interactions beyond seasonal timescales. These successful 3D global ocean simulations suggest that the model can be readily extended to other challenging components of the Earth system, marking a key milestone toward sustainable, data-driven climate prediction.

\section{Methods}\label{sec:methods_section}

\subsection{Generator}

The Korea Institute of Science and Technology’s Ocean model (KIST-Ocean) was developed based on a visual attention adversarial network composed of a generator and a discriminator\cite{Guo2023, Li2023}. The generator employs a U-shaped visual attention network (VAN) architecture, which reduces the dimensionality of feature maps via down-sampling and subsequently restores them through up-sampling. This design effectively captures multi-scale features and global context, while skip connections between the two paths mitigate information loss, ensuring robust performance even with relatively small datasets\cite{Ronneberger2015}.

At each time step t, the generator takes as input four three-dimensional (3D) oceanic variables (each with 15 vertical levels), 2 2D oceanic variables, and 6 2D surface boundary forcing variables, amounting to 68 input channels (\cref{edtab:ed_tab1}). The generator then predicts oceanic fields five days ahead for all 6 oceanic variables (totaling 62 output channels), excluding the boundary forcing variables. Both the input and output fields cover the global ocean domains extending from 0$^{\circ}$ to 360$^{\circ}$ longitude and from 79$^{\circ}$S to 80$^{\circ}$N latitude.

The generator architecture (\cref{edfig:ed_fig1}) begins with a stem layer that expands the number of channels via point-wise convolution. Subsequently, two alternating cycles of VAN stages and down-sampling blocks reduce the spatial dimensions of the feature map to one-quarter of its original size. This is followed by three rounds of alternating VAN stages and up-sampling blocks, and then two additional rounds, to fully restore the original spatial dimensionality. Finally, a head layer transforms the feature map into 62 output channels corresponding to oceanic conditions at t+5 days, applying a land-sea mask to zero out land grid points, ensuring that the loss calculation is unaffected by land region. Skip connections, implemented via feature concatenation, link the first VAN stage to fifth, and the second VAN stage to the fourth.

Each VAN stage (\cref{edfig:ed_fig1}A) followed the structure proposed by Wang et al.\cite{wang2018}, comprising batch normalization, point-wise convolution, Gaussian error linear unit~(GELU) activation, large-kernel attention~(LKA), point-wise convolution, batch normalization, another point-wise convolution, $3\times3$ depth-wise convolution, GELU activation, another point-wise convolution, and layer normalization. 
Two residual connections were included within each VAN stage, and circular padding was applied to all convolutional operations to maintain longitudinal continuity. 
The LKA operation decomposes a $K\times K$ convolution into a $(2d-1)\times(2d-1)$ depth-wise convolution, a $\lceil K/d\rceil\times\lceil K/d\rceil$ depth-wise dilated convolution (with dilation~$d$), and a point-wise convolution. In this study we set $K=5$ and $d=2$. As shown in \cref{edfig:ed_fig1}B, the LKA process can be expressed as:%

\[
\text{Attention}
  = \operatorname{Conv}_{1\times 1}\!
    \bigl(
      \operatorname{DW\!--\!D\!--\!Conv}\bigl(\operatorname{DW\!--\!Conv}(V)\bigr)
    \bigr)
\]

\[
\text{Output}
  = \text{Attention} \,\odot\, V
\]

where V is the input feature map, Attention is the attention map, and $\odot$ represents the Hadamard (element-wise) product. Here, DW Conv($\cdot$) and DW D Conv($\cdot$) indicate depth-wise and depth-wise dilated convolutions, respectively, while Conv$_{1\times1}$ ($\cdot$) denotes point-wise convolution.

A down-sampling block (\cref{edfig:ed_fig1}C) consisted of 2$\times$2 max-pooling operation followed by a point-wise convolution, halving the spatial dimensions of the input feature. Conversely, an up-sampling block (\cref{edfig:ed_fig1}D) employed a 2$\times$2 transposed convolution followed by a point-wise convolution, doubling the spatial dimensions. This architecture enables the extraction of multi-scale representations of oceanic variables, facilitating accurate global ocean predictions at five-day intervals.

\subsection{Adversarial Training and Discriminator Design}

Deep learning (DL)-based global weather prediction models commonly extend lead time by repeatedly feeding their outputs back as inputs in an auto-regressive manner, which may accumulate biases and lead to unrealistic distributional drift. Previous studies have addressed this issue by applying methods such as the Kullback-Leibler divergence loss or diffusion models to ensure that predicted distribution remain realistic\cite{Chen2024, Ling2024, Price2024}. Similarly, this study adopts a conditional generative adversarial network (GAN) to closely align predicted distribution with the actual distribution44.

In this study, the discriminator in KIST-Ocean was built based on the PatchGAN approach proposed by Isola et al.\cite{Isola2017}. Unlike traditional GAN discriminators, which outputs a single scalar value for an entire image, PatchGAN independently evaluates multiple image patchs, effectively capturing high-frequency local details and thereby enhancing the realism of generated fields.

As illustrated in \cref{fig:fig1}D, the KIST-Ocean discriminator receives both the predicted and ground truth fields at t+5 d as input. These inputs are processed through six 3$\times$3 depth-wise convolutional layers and interleaved with four 2$\times$2 max-pooling layers, applying GELU activations throughout except for the final convolution layer, which uses a sigmoid activation. The discriminator outputs a tensor with dimensions of 23$\times$10$\times$62, where each element represents the probability (ranging from 0 to 1) that the corresponding 16$^{\circ}$$\times$16$^{\circ}$ patch is real. This architecture effectively improves the alignment of local distribution and enables the capture of fine-scale oceanic features.

\subsection{Training Processes}

In this study, we employed transfer learning to address the limited availability of observational (reanalysis) training data. Transfer learning, which leverages knowledge from a preexisting model to accelerate adaptation to new tasks, has shown considerable success in climate prediction applications, including El Niño-Southern Oscillation (ENSO) forecasting\cite{Ham2019, Yosinski2014}. We pre-trained KIST-Ocean using two ensemble members from the CESM2-LE historical simulations spanning the period 1850--2014 (Extended Data Table 2)\cite{Rodgers2021}. Specifically, we utilized the entire period (1850--2014) from the first ensemble member (1301.012) and the period 1850--2004 from the second member (1301.013) for training, while reserving the period 2005--2014 from the second member for validation. This strategy ensured the incorporation of diverse historical simulations and maintained an independent validation dataset.

Following pretraining, the model was fine-tuned for 50 epochs using 2,336 reanalysis samples covering the period from 2014 to 2023, without additional validation. During pretraining, the model was trained for 100 epochs, and the version exhibiting the lowest objective validation loss was selected. All training procedures were conducted using a single NVIDIA A100 GPU, requiring approximately 33.3 h for pretraining and 2.4 h for fine-tuning, highlighting the computational efficiency of the proposed approach.

We employed PyTorch’s “ReduceLROnPlateau” for learning rate scheduling, decreasing the learning rate whenever improvements in the objective function plateaued. In the pretraining phase, the initial learning rate was set to 0.001, whereas it was set to 0.0001 during fine-tuning. Both phases employed the AdamW optimizer\cite{Loshchilov2017} with a weight decay of 0.1, a batch size of 20, and no dropout or droppath techniques.

\subsection{Inference Process}

KIST-Ocean employs an auto-regressive inference approach, repeatedly utilizing model output as input for subsequent prediction steps (\cref{fig:fig1}B), a method widely adopted in DL-based global prediction models\cite{Bi2023, Lam2023, Chen2024, Chen2023a, Chen2023b, Guo2024a, Bodnar2024}. In this study, we iterated this procedure 40 times to generate predictions spanning lead times of 5--200 d from the initial state. Because KIST-Ocean does not produce surface boundary conditions, these were prescribed either from ground truth (KIST-O\_GT) or climatology (KIST-O\_Clim). Climatology represents the minimal information typically available from a coupler, thus, the performances of KIST-O\_GT and KIST-O\_Clim effectively indicate the upper and lower bounds of predictive skill achievable by future DL-based coupled ocean-atmosphere models.

\subsection{Visual Attention Network}

The VAN employs a self-attention mechanism combined with large-kernel convolutions to effectively capture long-range dependencies and produce attention maps\cite{Guo2023}. Although large kernels generally require substantial computational resources and numerous parameters, VAN mitigates these limitations by decomposing convolutions through LKA, thereby achieving broader receptive fields with fewer parameters. Consequently, KIST-Ocean attains a receptive field size of $9\times9$ with only 6.6 million parameters, extending up to $36^{\circ} \times 36^{\circ}$ in VAN Stage 3, which is sufficient for resolving oceanic phenomena occurring on sub-weekly timescales.

\subsection{Partial Convolution}

Partial convolution, commonly employed in image inpainting tasks, excludes masked regions—often filled with zeros— from convolution operations, thus avoiding potential distortions. Instead, it performs convolutional operations solely on valid grid points and normalizes the outputs by the number of non-masked points\cite{Liu2018}. Owing to these advantages, partial convolution has been effectively applied to climate information reconstruction and assimilation of 3D global ocean temperature fields\cite{Ham2024, Kadow2020}. 
Specifically, for a feature map $V$ with a binary mask $M$ (defined as zeros and ones), the partial convolution PConv$(V,M)$ is computed as:

\begin{equation}
  \operatorname{PConv}(V,M)
    = \frac{\displaystyle \sum \nolimits_{i\in\Omega} w_i\,v_i \cdot m_i}
          {\displaystyle \sum \nolimits_{i\in\Omega} m_i + \varepsilon}
\end{equation}

where $\Omega$ represents the kernel region, $w_{i}$ denotes the convolution filter weights, $m_{i}=1$ identifies a valid grid point, and $m_{i}=0$ denotes a masked (invalid) grid point. The constant $\varepsilon$ is a small value added to prevent division by zero.

When processing oceanic variables, land grid cells are treated as masked regions, which would otherwise distort the learning of convolution kernels. To prevent such distortions, we replaced all convolution operations within the generator, except for point-wise convolutions, with partial convolutions.

\subsection{Objective functions}

In this study, the objective functions $L_{G}$ and $L_{D}$ for generator $G$ and discriminator $D$, which constitute KIST-Ocean, were defined as follows:

\begin{equation}
    \mathcal{L}_{G}
      = -\alpha \, \log D\!\bigl(G(X)\bigr)
        + \beta \,\mathcal{L}_{1}
\end{equation}

\begin{equation}
  \mathcal{L}_{D}
    = -\Bigl[\,\log D(Y)
            + \log\bigl(1 - D\!\bigl(G(X)\bigr)\bigr)
      \Bigr]
\end{equation}

where $X$ and $Y$ represent the input and ground truth, respectively, and are tensors with dimensions $N_{lon} \times N_{lat} \times N_{ch}$. $N_{lon}$, $N_{lat}$, and $N_{ch}$ denote the number of longitude and latitude grid points and the number of channels, respectively. The generator’s output $G(\cdot)$ and the discriminator’s output $D(\cdot)$ are tensors with dimensions $N_{lon} \times N_{lat} \times N_{ch}$ and $\lceil\frac{N_{lon}}{16}\rceil \times \lceil\frac{N_{lat}}{16}\rceil \times N_{ch}$, respectively.

In $L_{G}$, the balance between the adversarial loss term ($-\alpha \log D(G(X))$) and the reconstruction loss term ($\beta L_{1}$) is controlled by the regularization parameters $\alpha$ and $\beta$, respectively. We set $\alpha$ and $\beta$ to 0.4 and 0.6, respectively. That is, $L_{G}$ decreases as the discriminator classifies the generator’s output closer to one (i.e., the discriminator perceives the generator’s output as more realistic) and as the error between the generator’s output and the ground truth decreases. In contrast, $L_{D}$ decreases as the discriminator classifies the ground truth closer to one and the generator’s output closer to zero.

The reconstruction loss $L_{1}$ is defined as follows:

\begin{equation}
  \mathcal{L}_{1}
    = \frac{1}{N_{\mathrm{lon}}\,N_{\mathrm{lat}}\,N_{\mathrm{ch}}}
      \sum_{i=1}^{N_{\mathrm{lon}}} 
      \sum_{j=1}^{N_{\mathrm{lat}}} 
      \sum_{k=1}^{N_{\mathrm{ch}}}
        A(\phi_{j}) \,\Bigl|\,G(X)_{i,j,k} - Y_{i,j,k}\Bigr|
\end{equation}

where $A(\phi_{j})$ represents the latitude weight at latitude $\phi_{j}$ and is defined as $N_{\mathrm{lat}}\,\cos\phi_{j}\; / \sum_{j=1}^{N_{\mathrm{lat}}}\cos\phi_{j}$.

\subsection{Evaluation Metrics}

To assess global 3D ocean modeling performance, this study employed the root mean square error (RMSE) and anomaly correlation coefficient (ACC). For all metrics, the global average was calculated using the latitude-based weighting function $A(\phi_{j})$ defined in the “Objective Functions” section. Accordingly, the global mean RMSE between the ground truth $Y_{t,i,j,k}$ and forecast $\hat{Y}_{t,i,j,k,l}$ for channel $k$ and lead time $l$ is given by:

\begin{equation}
  \mathrm{RMSE}_{k,l} \;=\;
  \sqrt{\,
    \frac{1}{N_{\mathrm{time}}\,N_{\mathrm{lon}}\,N_{\mathrm{lat}}}
    \sum_{t=1}^{N_{\mathrm{time}}}
    \sum_{i=1}^{N_{\mathrm{lon}}}
    \sum_{j=1}^{N_{\mathrm{lat}}}
    A(\phi_{j})\,
    \bigl[
      (\hat{Y}_{t,i,j,k,l}-\overline{\hat{Y}}_{t,j,k,l})
        - (Y_{t,i,j,k}-\overline{Y}_{t,j,k})
    \bigr]^{2}
  }\,.
\end{equation}

where $N_{time}$, $N_{lon}$, and $N_{lat}$ denote the number of time samples, longitudes, and latitudes, respectively, and $\overline{(\cdot)}$ indicates the temporal climatology (from the 1st to the 73rd pentad).

Because Pearson’s correlation coefficient can be underestimated when directly averaged over an asymmetrically distributed sample\cite{Silver1987}, simply averaging grid-wise ACC values globally may introduce bias. To mitigate this issue, a Fisher’s Z-transform was applied to each grid point’s ACC; the transformed values were then averaged, and the result was finally inverse-transformed.

First, the ACC at each grid point $(i, j)$ for channel $k$ and lead time $l$ is defined as:

\begin{equation}
  \mathrm{ACC}_{i,j,k,l}
    \;=\;
    \frac{
      \displaystyle\sum\nolimits_{t=1}^{N_{\mathrm{time}}}
        \Bigl(\hat{Y}_{t,i,j,k,l}-\overline{\hat{Y}}_{t,\,\cdot,\,j,k,l}\Bigr)
        \Bigl(Y_{t,i,j,k}-\overline{Y}_{t,\,\cdot,\,j,k}\Bigr)
    }{
      \displaystyle
      \sqrt{
        \sum\nolimits_{t=1}^{N_{\mathrm{time}}}
          \Bigl(\hat{Y}_{t,i,j,k,l}-\overline{\hat{Y}}_{t,\,\cdot,\,j,k,l}\Bigr)^{2}
        \;\sum\nolimits_{t=1}^{N_{\mathrm{time}}}
          \Bigl(Y_{t,i,j,k}-\overline{Y}_{t,\,\cdot,\,j,k}\Bigr)^{2}
      }
    }
\end{equation}

The global mean ACC ($ACC_{k,l}$) was then computed by applying the Fisher’s Z-transform to each $ACC_{i,j,k,l}$, taking the average of these transformed values, and subsequently applying the inverse Z-transform:

\begin{equation}
  \mathrm{ACC}_{k,l}
    = \tanh\!\Biggl(
        \frac{1}{N_{\mathrm{lon}}\,N_{\mathrm{lat}}}
        \sum_{i=1}^{N_{\mathrm{lon}}}
        \sum_{j=1}^{N_{\mathrm{lat}}}
          A(\phi_{j})\,
          \operatorname{arctanh}\!\bigl(\mathrm{ACC}_{i,j,k,l}\bigr)
      \Biggr)
\end{equation}

where $tanh$ and $arctanh$ denote the hyperbolic tangent and inverse hyperbolic tangent functions, respectively.

\subsection{Idealized Wind Stress Forced Experiments}

In this study, we conducted three types of wind stress forcing experiments to evaluate, from multiple perspectives, whether DL-based KIST-Ocean can simulate physically consistent responses to atmospheric surface boundary forcing.

\paragraph{Westerly and Easterly Wind Bursts}
In the tropical Pacific, westerly wind bursts (WWBs) trigger eastward-propagating downwelling Kelvin waves, deepening the thermocline and warming SSTs, thereby amplifying El Niño conditions. Concurrently, westward-propagating upwelling Rossby waves are generated, some of which reflect off the Maritime Continent as eastward-propagating equatorial upwelling Kelvin waves. These reflected waves subsequently shoal the thermocline and cool SSTs, potentially mitigating El Niño intensity—a central mechanism described by the delayed oscillator theory of ENSO.

In this study, we conducted WWB and easterly wind burst (EWB) experiments by applying zonal wind stress anomalies in the western tropical Pacific (130$^{\circ}$--170$^{\circ}$E, 10$^{\circ}$S--10$^{\circ}$N) to verify whether KIST-Ocean reproduces the generation and propagation of Kelvin and Rossby waves in a manner consistent with that of the delayed oscillator theory. Specifically, we prescribed an idealized zonal wind stress forcing $\operatorname{frc}_{i,j}$ following a Gaussian distribution:

\begin{equation}
 \operatorname{frc}_{i,j}
   \;=\;
   \gamma\,
   \sin\!\Bigl(\frac{\pi(i+0.5)}{N_{\mathrm{lon}}}\Bigr)\,
   \sin\!\Bigl(\frac{\pi(j+0.5)}{N_{\mathrm{lat}}}\Bigr),
   \quad
   i\in\{0,1,\dots,N_{\mathrm{lon}}\},\;
   j\in\{0,1,\dots,N_{\mathrm{lat}}\}
\end{equation}

where $N_{lon}$ and $N_{lon}$ denote the numbers of grid points in the zonal and meridional directions of the forcing region, respectively. The scale parameter $\gamma$ (amplitude) was set to either 1 or -1, defining a WWB run ($\gamma=1$) and an EWB run ($\gamma=-1$). The initial ocean state was taken from December 14, 2013 (averaged over December 14--16), and climatological surface boundary conditions were prescribed. In the forcing region, the WWB and EWB runs added $\operatorname{frc}_{i,j}$ to the climatological zonal wind, but only during the first inference step, such that the subsequent wave responses could be traced to the imposed perturbation.

A control run using the same climatological boundary conditions, but without any additional perturbations, was performed for comparison. We examined 20$^{\circ}$C isotherm depth anomalies to investigate the generation and propagation of Kelvin and Rossby waves. Specifically, anomalies were calculated by subtracting the control run from the WWB and EWB runs, thereby isolating the wave responses triggered by WWBs and EWBs within KIST-Ocean.

\paragraph{Rossby Wave Propagation Experiments}
In the Pacific Ocean, Rossby waves generated by wind stress perturbations propagate at speeds that are inversely proportional to latitude and directly proportional to thermocline depth\cite{Chelton2007}. To assess whether KIST-Ocean accurately reproduces these fundamental relationships, we conducted WWB experiments at various latitudes in the tropical Pacific. In these experiments, the forcing region was fixed at 160$^{\circ}$--140$^{\circ}$W, while latitude bands were incrementally shifted from 0$^{\circ}$--5$^{\circ}$S to 5$^{\circ}$--10$^{\circ}$S in the Southern Hemisphere and from 0$^{\circ}$--5$^{\circ}$N to 5$^{\circ}$--10$^{\circ}$N in the Northern Hemisphere, resulting in 12 latitude bands ($N_{lon}=21$, \, $N_{lat}=6$; \cref{edtab:ed_tab3}).

We conducted a total of 432 WWB perturbation experiments (12 latitudes $\times$ 36 initial conditions), initialized from the central pentad of each month during 2014--2016 (e.g., January 13, February 17, and March 14), with WWB forcing applied only at the initial inference step. Control experiments were performed using identical initial conditions and prescribed climatological forcing without WWB perturbations. These experiments enabled assessment of KIST-Ocean’s capability to accurately simulate Rossby wave propagation and the latitude-dependent variability of wave speeds.

\paragraph{Rotational Wind Stress Forcing}
Strong cyclonic wind stress forcing, such as from tropical cyclones, induces Ekman transport, driving surface water radially outward and causing upwelling and cooling, whereas anticyclonic forcing leads to central warming. To evaluate if KIST-Ocean realistically reproduces these oceanic responses, we conducted experiments precribing cyclonic and anticyclonic wind stress forcing.

In this study, idealized rotational wind stress perturbations were defined using Gaussian functions. For a two-dimensional grid system, let $i_{0}=(N_{lon}+1) / 2$ and $j_{0}=(N_{lat}+1) / 2$ denote the zonal and meridional centers of the forcing region, respectively, and let $r_{i,j}=\sqrt{(i-i_{0})^{2}+(j-j_{0})^{2}}$ represent the distance of each grid point $(i, j)$ from the forcing center. The zonal $u_{i,j}$ and meridional $v_{i,j}$ components of the rotational wind stress perturbation are defined as follows:

\begin{equation}
u_{i,j}
  = \gamma\Bigl(-\frac{\,j-j_{0}\,}{r_{i,j}}\Bigr)\exp\!\Bigl(-\frac{r_{i,j}^{2}}{2\lambda^{2}}\Bigr),
\quad
v_{i,j}
  = \gamma\Bigl(-\frac{\,i-i_{0}\,}{r_{i,j}}\Bigr)\exp\!\Bigl(-\frac{r_{i,j}^{2}}{2\lambda^{2}}\Bigr)
\end{equation}

where the scale parameter $gamma$ determines the circulation type: $\gamma > 0$ corresponds to cyclonic circulation, while $\gamma < 0$ indicates anticyclonic circulation. The Gaussian spread $\gamma$ controls how rapidly (or gradually) the wind stress forcing amplitude decays from the center; in this experiment, $\lambda=1$.

The forcing region was set to 131$^{\circ}$--140$^{\circ}$E, 11$^{\circ}$--20$^{\circ}$N—a typical tropical cyclone formation zone ($N_{lon}=10$, \, $N_{lat}=10$). initial oceanic conditions were taken from June 17, 2015, with climatological surface boundary conditions prescribed. In the cyclone ($\gamma=1$) and anticyclone ($\gamma=-1$) experiments, persistent zonal and meridional wind stress perturbations were applied at each inference step, whereas the control run included no perturbations.

To assess the temperature response to rotational wind stress forcing, potential temperature anomalies at a depth of 105 m were calculated by subtracting the control run from the cyclone and anticyclone experiments. The resulting spatial anomaly patterns indicate how continuous rotational wind stress forcing modifies subsurface ocean temperature structures in KIST-Ocean.

\paragraph{KIST-Ocean Rossby Wave Propagation Speed}
In this study, we calculated the propagation speed of Rossby waves\cite{Killworth1997} using the following three-step procedure:

Step 1. Take the latitudinal mean of the 20$^{\circ}$C isotherm depth anomaly, $\mathrm{Z20}_{i,j,t}$ (m), for longitude $i$ ($^{\circ}$), latitude $j$ ($^{\circ}$), and lead time $t$ (days) over the Rossby wave detection zone (see \cref{edtab:ed_tab3}) and apply one-dimensional (1D) Gaussian filtering. By expressing this mathematically:

\begin{equation}
  [\mathrm{Z20}]_{i,t}
    = \Biggl(
        \frac{1}{\phi_{2}-\phi_{1}}
        \sum_{j=\phi_{1}}^{\phi_{2}} \mathrm{Z20}_{i,j,t}
      \Biggr)
      * \frac{1}{\sqrt{2\pi\sigma}}\,
      \exp\!\Bigl( -\tfrac{i^{2}}{2\sigma^{2}} \Bigr)
\end{equation}

where $\phi_{1}$ and $\phi_{2}$ represent the latitudes of the equatorward and poleward edges of the Rossby wave detection zone, respectively ($^{\circ}$). We defined the Rossby wave detection zone as the region extending from 1.5$^{\circ}$ ($\phi_{1}$) to 5.5$^{\circ}$ ($\phi_{2}$) poleward of the central latitude of the wind stress forcing (\cref{edtab:ed_tab3}). The symbol * denotes the zonal convolution operator (i.e., 1D convolution), while $\sigma$ is a parameter that controls the intensity of the Gaussian blur (here, $\sigma = 9$).

Step 2. Find the longitude $i_{t}^{*}$, at which $[\mathrm{Z20}]_{i,t}$ reaches its minimum (unit: $^{\circ}$).

\begin{equation}
  i_{t}^{*} \;=\; \operatorname*{arg\,min}_{i}\,([\mathrm{Z20}]_{i,t})
\end{equation}

Step 3. Finally, the Rossby wave propagation speed $C_{Z}$ (unit: $m s^{-1}$) was calculated as follows:

\begin{equation}
  C_{Z}
    = \frac{\operatorname{Rad}\!\bigl(i_{t=55}^{*}-i_{t=5}^{*}\bigr)\,R}
          {50 \times 86400}
\end{equation}

where R represents the Earth’s radius, defined in this study as 6,378,000 m. The denominator 50$\times$86400 converts the time unit from 50 days to seconds.

\paragraph{Rossby Wave Theoretical Phase Speed}
To determine whether KIST-Ocean realistically reproduces Rossby wave propagation speeds, we estimated theoretical phase speeds assuming a simplified first baroclinic Rossby wave under latitude and thermocline depth conditions, employing long-wave approximation. However, real oceanic conditions—such as background currents, nonlinear mixing, and spatially variable wind stress—can introduce differences between theory and observations\cite{Chelton1996}. Assuming sufficiently large wavelengths, the phase speed $\operatorname{C}_{Phase}$ of the first baroclinic Rossby wave can be approximated by\cite{Killworth1997}:

\begin{equation}
  C_{\text{Phase}} \;=\; -\beta \,\frac{c_{0}^{2}}{f^{2}}
\end{equation}

where the negative sign indicates westward propagation, $C_{0} = \sqrt{g' H}$ is the first internal gravity wave speed, $f = 2\Omega\cos\phi$ is the Coriolis parameter at latitude $\phi$, and $\beta = \partial f / \partial y \approx 2\Omega\cos\phi / R_{E}$ is the beta term. Here, $\Omega \approx 0.03\;\mathrm{s^{-1}}$ is Earth’s angular rotation speed, $R_{E} \approx 6.37\times10^{6}\;\mathrm{m}$ is Earth’s radius, $g' \approx 0.03\;\mathrm{m\,s^{-2}}$ is the reduced gravity, and $H$ is the thermocline depth. Thus, as latitude increases, the phase speed decreases, and as the thermocline deepens, the phase speed increases.

\subsection{FourCastNet-Based Model}

We employed FourCastNet, an adaptive Fourier neural operator (AFNO)-based global weather prediction model\cite{Guibas2021}, to investigate the sensitivity of global 3D ocean forecasts to different DL algorithms. Originally trained on the ERA5 dataset (1979--2018, 6-hour intervals, 0.25$^{\circ}$ resolution, 20 atmospheric variables)\cite{Kurth2023}, FourCastNet provides predictions up to 10 days. In this study, we configured FourCastNet using a 4$\times$4 patch size, embedding vectors with 500 dimensions, and 5 AFNO blocks, resulting in 25 million parameters—nearly 4 times more than KIST-Ocean. The model was trained following the same two-phase approach as KIST-Ocean, applying a learning rate of 0.001 during pretraining (CESM2-LE) and 0.0001 during fine-tuning (reanalysis data), with optimization performed using the FuseAdam optimizer.

\section*{Data availability}
\label{section:datasets}

The datasets used in this study can be downloaded from the following sources: CESM2-LE, https://doi.org/10.26024/kgmp-c556; GODAS, https://psl.noaa.gov/data/gridded/data.godas.html; OISST v.2, https://psl.noaa.gov/data/gridded/data.noaa.oisst.v2.highres.html; ERA5, https://doi.org/10.24381/cds.adbb2d47; and NMME hindcasts and real-time forecasts, https://iridl.ldeo.columbia.edu/SOURCES/.Models/.NMME/.

\section*{Code availability}
PyTorch (https://pytorch.org) libraries were used to formulate a deep learning-based global ocean model (KIST-Ocean). The Source code for FourCastNet is available at https://github.com/NVlabs/FourCastNet.

\section*{Acknowledgments}
Funding: This study was supported by the Korea Institute of Science and Technology (grant number 2E33621). J.-H. Kim was supported by the National Research Foundation of Korea (NRF) grant funded by the Korean Government Ministry of Science and ICT (MSIT) RS-2025-00553285. D. Kang was supported by NRF grants funded by the Korean government (NRF-2021R1C1C2004621 and NRF-2022M3K3A1094114) and the Korea Meteorological Administration Research and Development Program (RS-2025-02313090).

\section*{Author Contributions}
J.-H. K. and D. K. designed the study. J.-H. K., D. K., Y.-M. Y., J.-H. P., and Y.-G. H. contributed to writing the manuscript. J.-H. K. and D. K. performed the experiments and analyses. All authors discussed the results and contributed significantly to the manuscript.

\section*{Competing Interests}
The authors declare no competing financial or non-financial interests.

\section*{Corresponding Author}
Correspondence to Jeong-Hwan Kim: jeonghwan@kist.re.kr, and Correspondence to Daehyun Kang: dkang@kist.re.kr

\clearpage
\backmatter

\appendix  
\section*{Extended data}

\setcounter{table}{0}
\renewcommand{\thetable}{\arabic{table}}
\renewcommand{\tablename}{Extended Data Table}

\setcounter{figure}{0}
\renewcommand{\thefigure}{\arabic{figure}}
\renewcommand{\figurename}{Extended Data Figure}

\begin{edtable}[h!]
  \centering
  \caption{\textbf{Ocean variables and surface boundary forcing variables used in the Korea Institute of Science and Technology’s ocean model (KIST-Ocean).}}
  \label{edtab:ed_tab1}
  \begin{tabular}{lll}
  \toprule
  \textbf{Variable} & \textbf{Layer} & \textbf{Description (unit)} \\
  \midrule
  \multicolumn{3}{l}{\textit{Ocean}} \\
  $\theta$     & 15    & Potential temperature ($^{\circ}$C)     \\
  UO           & 15    & Zonal current ($m \, s^{-1}$) \\
  VO           & 15    & Meridional current ($m \, s^{-1}$) \\
  S            & 15    & Salinity ($kg \, kg^{-1}$)\\
  SST          & 1     & Sea surface temperature ($^{\circ}$C)\\
  SIC          & 1     & Sea ice concentration (\%)\\
  \midrule
  \multicolumn{3}{l}{\textit{Forcing}} \\
  $\tau_{x}$   & 1     & Zonal wind stress ($N \, m^{-2}$)\\
  $\tau_{y}$   & 1     & Meridional wind stress ($N \, m^{-2}$)\\
  LW           & 1     & Downward long-wave radiation ($W \, m^{-2}$)\\
  SW           & 1     & Downward shortwave radiation ($W \, m^{-2}$)\\
  LHF          & 1     & Latend heat flux ($W \, m^{-2}$)\\
  SHF          & 1     & Sensible heat flux ($W \, m^{-2}$)\\
  \bottomrule
  \end{tabular}
  \end{edtable}

  \begin{edtable}[ht]
    \centering
    \caption{\textbf{Datasets used for training and forecast skill evaluation of KIST-Ocean, including their periods and total sample counts.} Note that “1301.012” and “1301.013” denote the ensemble member identification numbers from the CESM2-LE project.}
    \label{edtab:ed_tab2}
    \begin{tabular}{llll}
    \toprule
      & \textbf{Data sources} & \textbf{Periods} & \textbf{No.\ of samples} \\
    \midrule
    \multirow{3}{*}{\textbf{Pretraining}}
      & CESM2--LE & 1850--2014 for 1301.012 & 23,360 \\
      &           & 1850--2004 for 1301.013 &         \\
      & CESM2--LE & 2005--2014 for 1301.013 & \hphantom{0,}730 \\[0.6em]
    \multirow{3}{*}{\textbf{Fine-tuning}}
      & GODAS & \multirow{3}{*}{1982--2013} &         \\
      & OISST &                                &  2,336       \\ 
      & ERA5  &                                &  \\ [0.6em]
    \multirow{3}{*}{\textbf{Testing}}
      & GODAS & \multirow{3}{*}{2014--2023} &         \\
      & OISST &                               & \hphantom{0,}730    \\
      & ERA5  &                               &  \\
    \bottomrule
    \end{tabular}
    \end{edtable}

  \begin{edtable}[h!]
    \centering
    \caption{\textbf{Latitude center of westerly wind stress forcing, forcing region, and latitude range of the Rossby wave detection zone for each nudging experiment.}}
    \label{edtab:ed_tab3}
    \begin{tabular}{lll}
    \toprule
    \textbf{Latitudinal center of forcing} & \textbf{Forcing region} & \textbf{Rossby wave detection zone} \\
    \midrule
    7.5$^{\circ}$S  &  160$^{\circ}$--140$^{\circ}$W, 5$^{\circ}$--10$^{\circ}$S  &  9$^{\circ}$--13$^{\circ}$S    \\
    6.5$^{\circ}$S  &  160$^{\circ}$--140$^{\circ}$W, 4$^{\circ}$--9$^{\circ}$S   &  8$^{\circ}$--12$^{\circ}$S    \\
    5.5$^{\circ}$S  &  160$^{\circ}$--140$^{\circ}$W, 3$^{\circ}$--8$^{\circ}$S   &  7$^{\circ}$--11$^{\circ}$S    \\
    4.5$^{\circ}$S  &  160$^{\circ}$--140$^{\circ}$W, 2$^{\circ}$--7$^{\circ}$S   &  6$^{\circ}$--10$^{\circ}$S    \\
    3.5$^{\circ}$S  &  160$^{\circ}$--140$^{\circ}$W, 1$^{\circ}$--6$^{\circ}$S   &  5$^{\circ}$--9$^{\circ}$S     \\
    2.5$^{\circ}$S  &  160$^{\circ}$--140$^{\circ}$W, 0$^{\circ}$--5$^{\circ}$S   &  4$^{\circ}$--8$^{\circ}$S     \\
    2.5$^{\circ}$N  &  160$^{\circ}$--140$^{\circ}$W, 0$^{\circ}$--5$^{\circ}$N   &  4$^{\circ}$--8$^{\circ}$S     \\
    3.5$^{\circ}$N  &  160$^{\circ}$--140$^{\circ}$W, 1$^{\circ}$--6$^{\circ}$N   &  5$^{\circ}$--9$^{\circ}$S     \\
    4.5$^{\circ}$N  &  160$^{\circ}$--140$^{\circ}$W, 2$^{\circ}$--7$^{\circ}$N   &  6$^{\circ}$--10$^{\circ}$S    \\
    5.5$^{\circ}$N  &  160$^{\circ}$--140$^{\circ}$W, 3$^{\circ}$--8$^{\circ}$N   &  7$^{\circ}$--11$^{\circ}$S    \\
    6.5$^{\circ}$N  &  160$^{\circ}$--140$^{\circ}$W, 4$^{\circ}$--9$^{\circ}$N   &  8$^{\circ}$--12$^{\circ}$S    \\
    7.5$^{\circ}$N  &  160$^{\circ}$--140$^{\circ}$W, 5$^{\circ}$--10$^{\circ}$N  &  9$^{\circ}$--13$^{\circ}$S    \\
    \bottomrule
    \end{tabular}
    \end{edtable}

\begin{edfigure}[h!]
	\centering
	\includegraphics[width=0.9\textwidth]{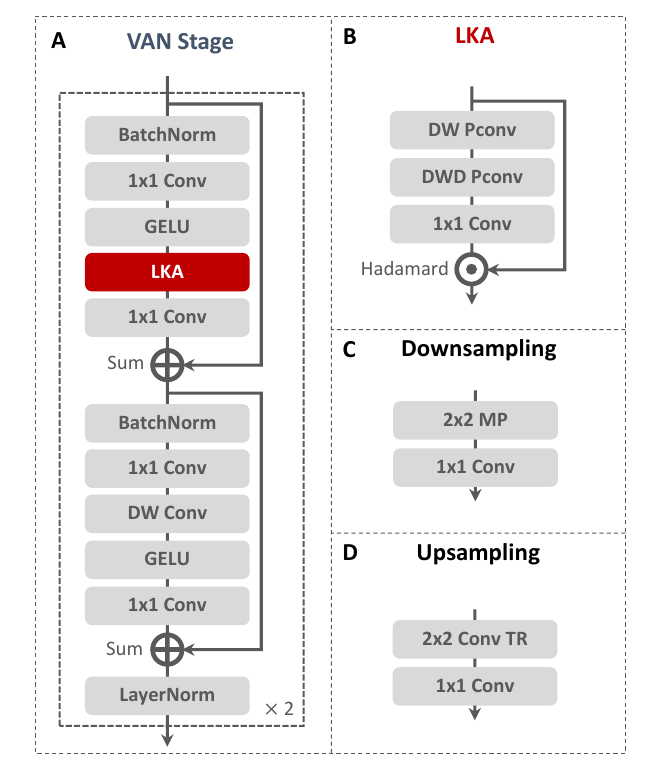}
	\caption{\textbf{Structures of the individual modules comprising the generator. A}, Architecture of the visual attention network (VAN) stage, where BatchNorm and LayerNorm refer to batch and layer normalization, respectively, and 1$\times$1 Conv and DW Conv denote point- and depth-wise convolutions, respectively. GELU refers to the Gaussian Error Linear Unit activation function. \textbf{B} Structure of the large kernel attention (LKA) module, in which DW Pconv and DWD Pconv represent depth-wise partial convolution and depth-wise dilated partial convolution, respectively. \textbf{C}, Down-sampling module, where MP indicates max pooling that halves the x and y dimensions via a 2$\times$2 window. D, Up-sampling module, featuring Conv transposed convolution (TR) to double the x and y dimensions through a 2$\times$2 operation.}
	\label{edfig:ed_fig1}
\end{edfigure}

\begin{edfigure}[h!]
	\centering
	\includegraphics[width=0.9\textwidth]{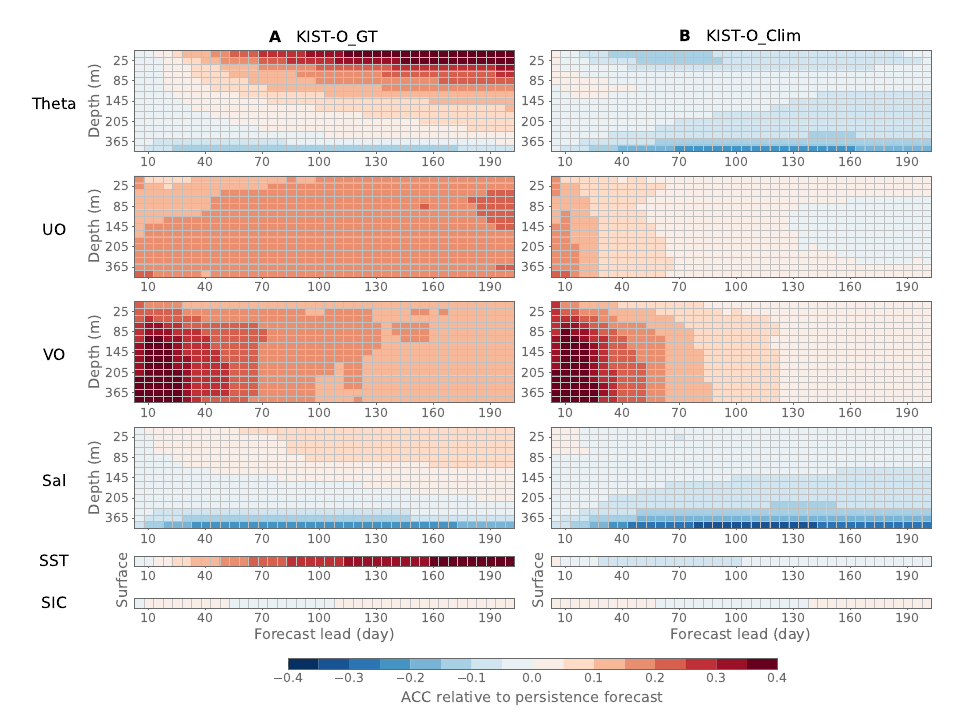}
	\caption{\textbf{Scorecard of relative anomaly correlation coefficient (ACC), computed over 2014--2023.} Here, “relative ACC” is defined as the difference between the ACC of each KIST-Ocean inference setup and that of the persistence forecast. The left column displays the relative ACC for KIST-O\_GT (with the ground truth prescribed as the surface boundary forcing), whereas the right column shows that for KIST-O\_Clim (with a climatology prescribed). From top to bottom, each row presents the relative ACC for potential temperature (Theta), zonal current (UO), meridional current (VO), salinity (Sal), sea surface temperature (SST), and sea ice concentration (SIC), respectively.}
	\label{edfig:ed_fig2}
\end{edfigure}

\begin{edfigure}[h!]
	\centering
	\includegraphics[width=0.9\textwidth]{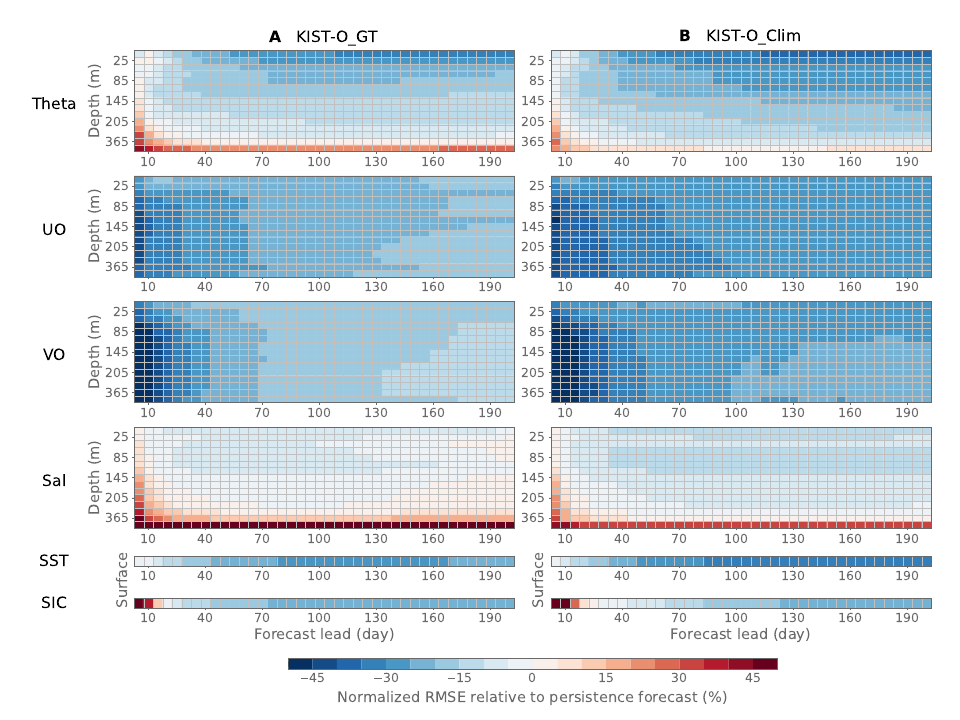}
	\caption{\textbf{Scorecard of normalized root mean square error (RMSE, \%) for each KIST-Ocean inference setup relative to the persistence forecast, computed over 2014--2023.} These scores are expressed as the percentage difference between the RMSE of each KIST-Ocean inference setup and that of the persistence forecast. The left column displays the normalized RMSE for KIST-O\_GT, whereas the right column shows that for KIST-O\_Clim. From top to bottom, each row presents the normalized RMSE for potential temperature (Theta), zonal current (UO), meridional current (VO), salinity (Sal), SST, and SIC, respectively.}
	\label{edfig:ed_fig3}
\end{edfigure}

\begin{edfigure}[h!]
	\centering
	\includegraphics[width=0.9\textwidth]{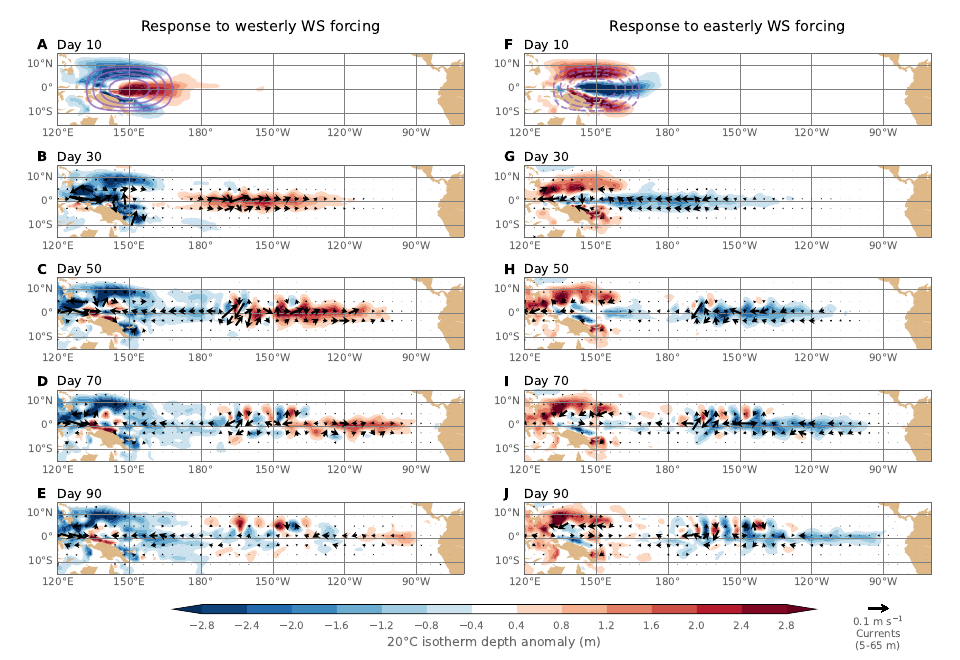}
	\caption{\textbf{Wave responses in the tropical Pacific to zonal wind stress forcing, as simulated by KIST-Ocean.} The left column shows responses to westerly wind stress, while the right column presents responses to easterly wind stress. From top to bottom, each panel corresponds to forecast lead times of 10, 30, 50, 70, and 90 d. Shading denotes the 20$^{\circ}$C isotherm depth anomaly (m); purple contours represent normalized zonal wind stress forcing (unitless); and black arrows denote depth-averaged (5--65 m) horizontal currents (m s$^{-1}$). The initial ocean state corresponds to December 12--16, 2013. For each experiment, a Gaussian-distributed forcing was nudged into the initial condition over 130$^{\circ}$--170$^{\circ}$E and 10$^{\circ}$S--10$^{\circ}$N (see “Idealized wind stress forcing nudging experiments” for details). Maps were generated using the Basemap Toolkit (v1.2.0).}
	\label{edfig:ed_fig4}
\end{edfigure}

\begin{edfigure}[h!]
	\centering
	\includegraphics[width=0.9\textwidth]{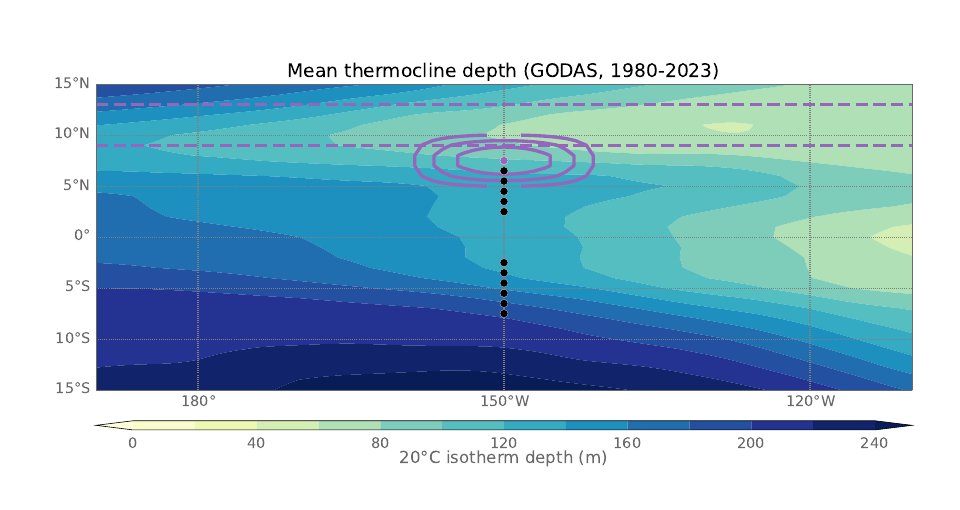}
	\caption{\textbf{Averaged thermocline depth and layout of westerly wind burst nudging experiments.} Shading shows the averaged 20 $^{\circ}$C isotherm depth over 1980--2023 (m). Dots represent the central locations of each nudging experiment (the purple dot indicates the northernmost experiment). Purple contours denote the normalized westerly wind forcing for the northernmost experiment (interval = 0.5; unitless), and the purple horizontal dashed lines indicate the Rossby wave detection zone for that experiment.}
	\label{edfig:ed_fig5}
\end{edfigure}

\begin{edfigure}[h!]
	\centering
	\includegraphics[width=0.9\textwidth]{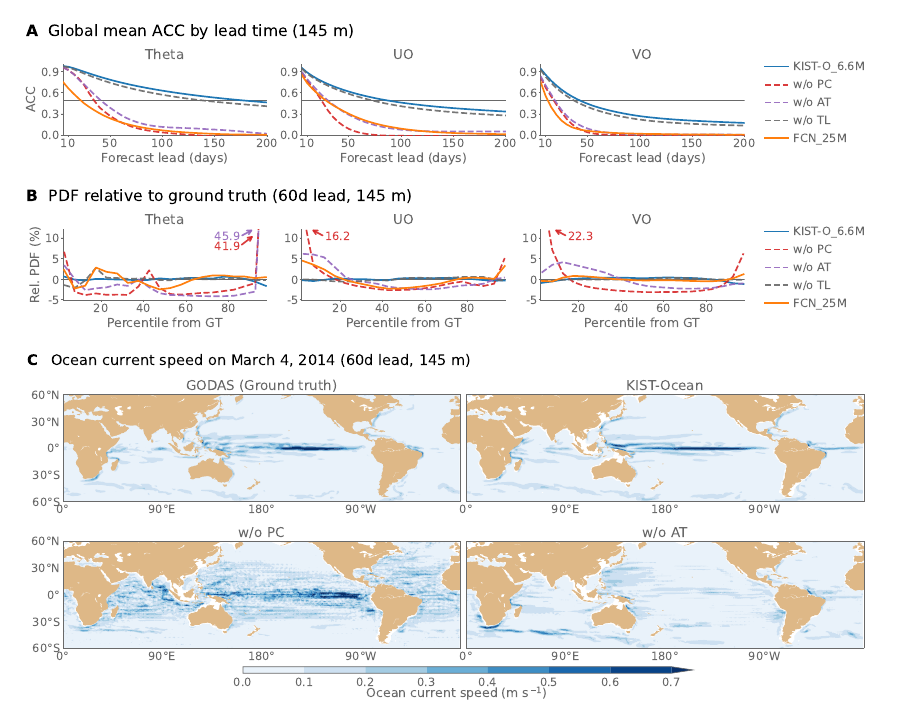}
	\caption{\textbf{Sensitivity experiments demonstrating how key algorithms affect KIST-Ocean’s predictive performance. A}, Globally averaged ACC for potential temperature (Theta), zonal current (UO) at a depth of 145 m, and meridional current (VO) at a depth of 145 m, as a function of forecast lead time over 2014--2023. KIST-O\_6.6M (blue solid line) is the original model; “w/o PC” (dashed red) excludes partial convolution; “w/o AT” (dashed purple) excludes adversarial training; “w/o TL” (dashed gray) excludes transfer learning (w/o TL); “FCN\_25M” (solid orange) is the FourCastNet-based model with 25 million parameters (for details, see the section “FourCastNet-Based Model” in Methods). The x-axis represents forecast lead time (days), and the y-axis represents ACC. \textbf{B}, Relative probability density functions (PDFs) comparing 60-day predictions to ground truth; values closer to zero indicate better alignment with observations. The x-axis represents percentiles based on the ground truth, and the y-axis represents the relative PDF (\%). \textbf{C}, Global ocean current speed at a depth of 145 m on March 4, 2024, with a 60-day forecast lead time. From left to right: GODAS (ground truth), KIST-O\_GT, the model without partial convolution ("w/o PC”), and the model without adversarial training (“w/o AT”). Maps were generated using the Basemap Toolkit (v1.2.0).}
	\label{edfig:ed_fig6}
\end{edfigure}

\begin{edfigure}[h!]
	\centering
	\includegraphics[width=0.9\textwidth]{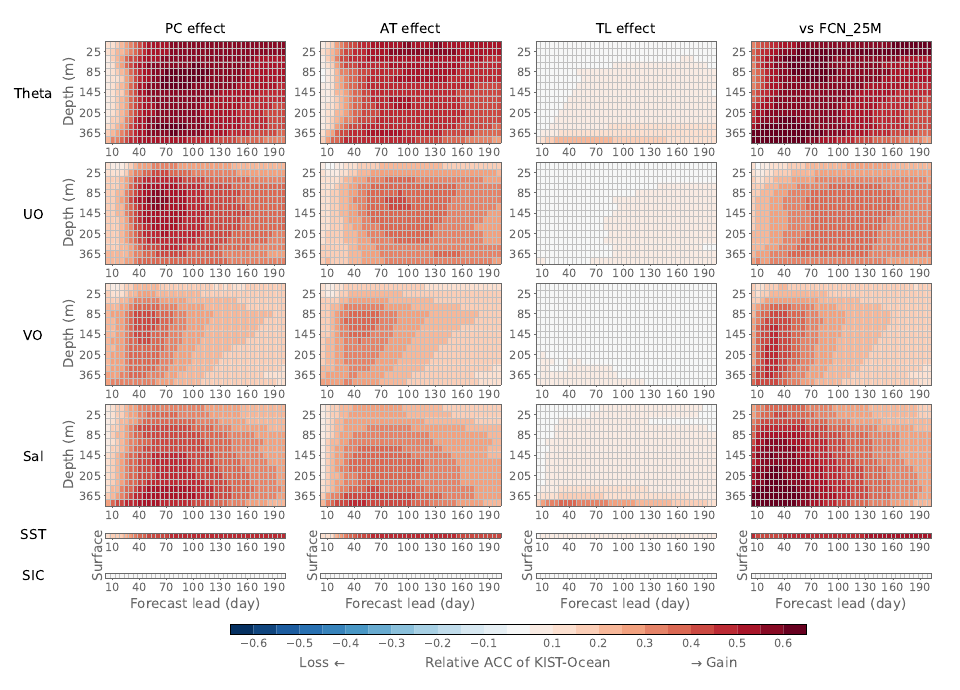}
	\caption{\textbf{Scorecard of differences in ACC between the original version of KIST-Ocean and versions with key algorithms removed, computed over the period 2014--2023.} From left to right, each column corresponds to the model without the partial convolution (PC effect), without adversarial training (AT effect), without transfer learning (TL effect), and the model based on FourCastNet with 25 million parameters (FCN\_25M), respectively. From top to bottom, each row shows the ACC for potential temperature (Theta), zonal current (UO), meridional current (VO), salinity (Sal), SST, and SIC, respectively.}
	\label{edfig:ed_fig7}
\end{edfigure}

\DeactivateWarningFilters[hyperreflevel]

\clearpage

\appendix 
\section*{Supplementary Information}\label{sec:supplementary}

\setcounter{table}{0}
\renewcommand{\thetable}{S\arabic{table}}
\renewcommand{\tablename}{Table S}

\setcounter{figure}{0}
\renewcommand{\thefigure}{S\arabic{figure}}
\renewcommand{\figurename}{Figure S}

\setcounter{section}{0}    
\renewcommand{\thesection}{Text~S\arabic{section}}

\crefname{table}{Table S}{Tables S}
\crefname{figure}{Fig. S}{Figs. S}
\crefname{section}{}{}

\section{Datasets}\label{sec:text_s1}

\subsection{CESM2 Large-Ensemble Simulations}

For pretraining of KIST-Ocean, we utilized two ensemble members (1301.012 and 1301.013) from the historical simulations (1850--2014) of the CESM2 Large Ensemble Community Project (CESM2-LE)\cite{Rodgers2021}. Model outputs were averaged over five-day intervals and interpolated horizontally onto a regular 1°×1° grid spanning 0°--360° longitude and 79°S--80°N latitude, and vertically onto 15 levels (depths of 5, 25, 45, 65, 85, 105, 125, 145, 165, 185, 205, 225, 265, 365, and 600 m) to facilitate transfer learning. This procedure yielded a total of 24,090 samples, of which 23,360 (320 years × 73 pentads) were used for training, while the remaining 730 (10 years × 73 pentads) were reserved for validation. Each variable was standardized using the z-score method, without removing climatological means (i.e., seasonal cycles).

\subsection{Reanalysis Dataset}
To fine-tune KIST-Ocean and assess its global 3D ocean performance, we employed several reanalysis datasets. 3D oceanic variables (potential temperature, zonal and meridional currents, and salinity) and surface wind stress were obtained from the National Centers for Environmental Prediction (NCEP) Global Ocean Data Assimilation System (GODAS)\cite{Behringer1998}. Two-dimensional oceanic variables (sea surface temperature (SST) and sea ice concentration) were derived from the National Oceanic and Atmospheric Administration Optimum Interpolation Sea Surface Temperature (OISST) version 2\cite{Huang2021}. Surface energy flux variables (downward shortwave and long-wave radiation, sensible and latent heat fluxes) were sourced from the European Centre for Medium-Range Weather Forecasts Reanalysis version 5 (ERA5)\cite{Hersbach2020} for the period 1982-2023.

For consistency, these data were interpolated onto the same horizontal 1°×1° grid and vertical levels utilized during pretraining. From the resulting total of 3,066 samples, 2,336 (covering the period 1982--2013) were allocated for fine-tuning, while the remaining 730 samples (covering 2014--2023) were reserved for evaluation. All variable were standardized using the z-score method without removing climatological means.

\subsection{The North American Multi-Model Ensemble (NMME) Forecasts}
To evaluate the SST predictions produced by KIST-Ocean, we utilized global SST forecasts from the NMME project provided by the International Research Institute34. Specifically, we selected the models COLA-RSMAS-CCSM4, GFDL-SPEAR, NASA-GEOSS2S, NCEP-CFSv2, and CanSIPS-IC3 for the verification period spanning 2014--2023 (\cref{sitab:si_tab1}). Each model, configured as a fully coupled atmosphere-ocean general circulation model, contributed 9--12 ensemble members CanSIPS-IC4 was excluded due to incomplete data availability. Monthly forecast anomalies were computed by removing for 2014--2023 separately at each forecast lead time.

\section{Comparison experiments of KIST-Ocean responses to two different surface boundary forcings}\label{sec:text_s2}
During boreal spring 2014, a strong westerly wind burst (WWB) over the western equatorial Pacific (140°--170°E), generated a pronounced downwelling Kelvin wave that propagated eastward and recharged the upper-ocean heat content to levels comparable with those of 1997. Despite this recharge, an El Niño failed to developed in the 2014/15 winter because easterly wind bursts (EWBs) in June 2014 produced an upwelling Kelvin wave that largely offset the earlier warming and inhibited activation of the Bjerknes feedback1,2. Consequently, the warm water volume persisted into early 2015. Between March and July 2015 a sequence of WWBs, both more frequent and markedly stronger than those of 2014, triggered the Bjerknes feedback and ultimately led to the strong 2015/16 El Niño3,4 (\cref{sifig:si_fig1}). This sequence offered an ideal natural experiment for testing whether KIST-Ocean could reproduce the coupled atmosphere-ocean processes that govern ENSO evolution.

To examine whether KIST-Ocean reproduces the oceanic response to such atmospheric forcing, we conducted two hindcast suites initialized on 14 December 2013 (EXP\_Dec2013) and 3 May 2015 (EXP\_May2015). For each initial state we ran (1) KIST-O\_GT, driven by observed surface boundary conditions, and (2) KIST-O\_Clim. These experiments allow us to assess the model’s physical consistency under both ideal and historically realized surface boundary forcing.

In EXP\_Dec2013, KIST-O\_GT successfully produced and propagated the downwelling Kelvin wave associated with the observed WWB and captured the subsequent rise in eastern Pacific sea surface temperature (SST) after April 2014 (\cref{sifig:si_fig2}A--C). Under climatological forcing, neither the Kelvin wave nor the SST warming emerged (\cref{sifig:si_fig2}D--F), a contrast that is highlighted in longitude-time Hovmöller diagrams (\cref{sifig:si_fig3}).

EXP\_May2015 illustrates the model’s behaviour when the equatorial Pacific thermocline is deeply recharged. With realistic WWBs applied, KIST-O\_GT closely followed the observed growth of the Nino3.4 anomaly and reproduced the transition to a strong El Niño after October 2015 (\cref{sifig:si_fig4}A). Note that, the Nino3.4 index is defined by averaging the SST over the region 170°--120°W and 5°S--5°N. In KIST-O\_Clim, the absence of WWB-related westerly stress allowed strong easterly currents to develop, rapidly discharging the stored heat and driving the system toward La Niña-like conditions (\cref{sifig:si_fig4}E--G). Longitude-depth sections reveal that, in KIST-O\_GT, convergence of warm surface water and the core of positive subsurface temperature anomalies gradually migrate from the eastern to the central Pacific, generating a pronounced warm center in the Nino3.4 region (\cref{sifig:si_fig4}B--D). Without westerly forcing, KIST-O\_Clim instead exhibits the rapid shoaling of the thermocline and a discharge phase typical of a transition from El Niño to La Niña.

Taken together, these hindcast experiments demonstrate that KIST-Ocean can reproduce key phenomena—Kelvin wave dynamics and ENSO evolution—over lead times exceeding several months when accurate surface forcing is supplied, confirming the model’s strong ocean simulation capability. The stark divergence between KIST-O\_GT and KIST-O\_Clim in EXP\_May2015 confirms that the model has learned the dynamical linkage between surface wind stress, pressure gradient forcing, and subsurface heat content, underscoring its ability to represent the coupled processes essential to ENSO evolution.

\section{Estimation of vertical velocity}\label{sec:text_s3}
Because KIST-Ocean does not explicitly predict ocean vertical velocity, w, we infer it diagnostically from the modelled horizontal velocities, $u$ (zonal) and $v$ (meridional), by integrating the three-dimensional continuity equation,

\begin{equation}
  \frac{\partial u}{\partial x}
  +\frac{\partial v}{\partial y}
  +\frac{\partial w}{\partial z}
  = 0
  \tag{S1}
  \label{eq:eq_s1}
  \end{equation}

Horizontal derivatives $\partial u / \partial x$ and $\partial v / \partial y$ are evaluated with centered finite differences on the model grid. Integrating Eq. (S1) vertically from the surface ($z=0$) to an arbitrary depth z and imposing a rigid-lid upper boundary condition $w(z=0)=0$, yields

\begin{equation}
  w(z) \;=\;
  - \int_{0}^{z} \!\left(
        \frac{\partial u}{\partial x}
        + \frac{\partial v}{\partial y}
      \right)\,dz'
  \tag{S2}
  \label{eq:eq_s2}
  \end{equation}

Equation (S2) provides a consistent estimate of vertical velocity throughout the water column based solely on the horizontal velocity fields produced by KIST-Ocean5.

\clearpage
\backmatter

\begin{sitable}[h!]
  \centering
  \caption{\textbf{Information on the five models from the North American Multi-Model Ensemble (NMME) project used for sea surface temperature (SST) prediction performance comparison with KIST-Ocean.} The second column indicates the period for each model as provided by International Research Institute (IRI), and the third column shows the number of ensemble members for each model.}
  \label{sitab:si_tab1}
  \begin{tabular}{lll}
  \toprule
  \textbf{Model} & \textbf{Period} & \textbf{Number of Ens.} \\
  \midrule
  COLA-RSMAS-CCSM4  &  Jan 1982--current   &  12    \\
  CanSIPS-IC3       &  Jan 1980--current   &  12    \\
  GFDL-SPEAR        &  Jan 1991--current   &  12    \\
  NASA-GEOSS2S      &  Feb 1982--current   &  9     \\
  NCEP-CFSv2        &  Jan 1982--current   &  10    \\
  \bottomrule
  \end{tabular}
  \end{sitable}

\begin{sifigure}[h!]
  \centering
  \includegraphics[width=0.9\textwidth]{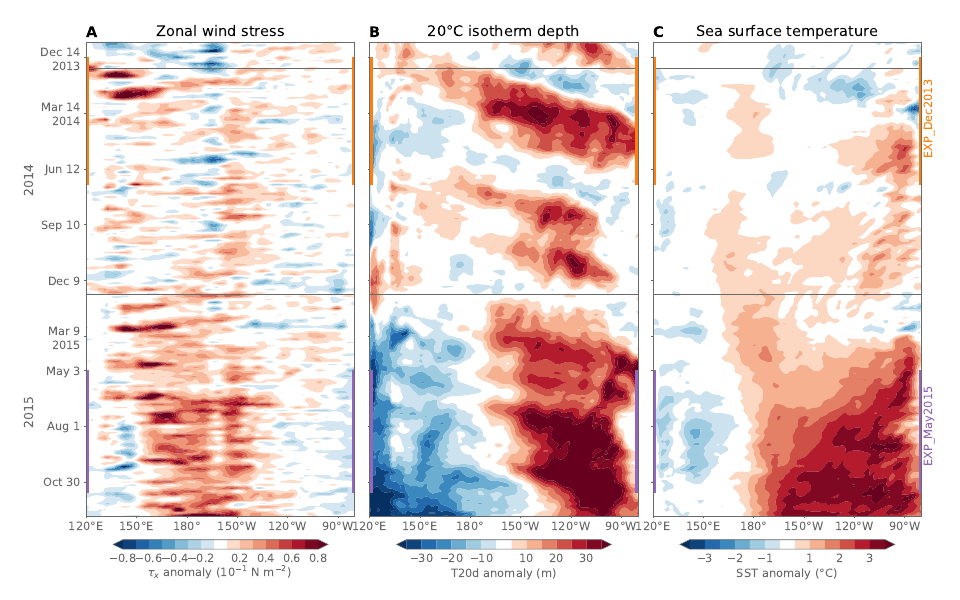}
  \caption{\textbf{Longitude-time evolution of equatorial Pacific conditions (19 November 2013--29 December 2015).} Hovmöller diagrams along the equator (2°S--2°N) showing anomalies of A, zonal wind stress (N m$^{-2}$), B, 20 °C isotherm depth (m; positive = thermocline deepening), and C, sea surface temperature (°C). All anomalies are computed with respect to the corresponding climatology. The orange and purple vertical bars denote the lead time ranges of two experiments generated with the KIST-Ocean, initiated from 14 December 2013 (EXP\_Dec2013) and 3 May 2015 (EXP\_May2015), respectively.}
  \label{sifig:si_fig1}
\end{sifigure}

\begin{sifigure}[h!]
  \centering
  \includegraphics[width=0.9\textwidth]{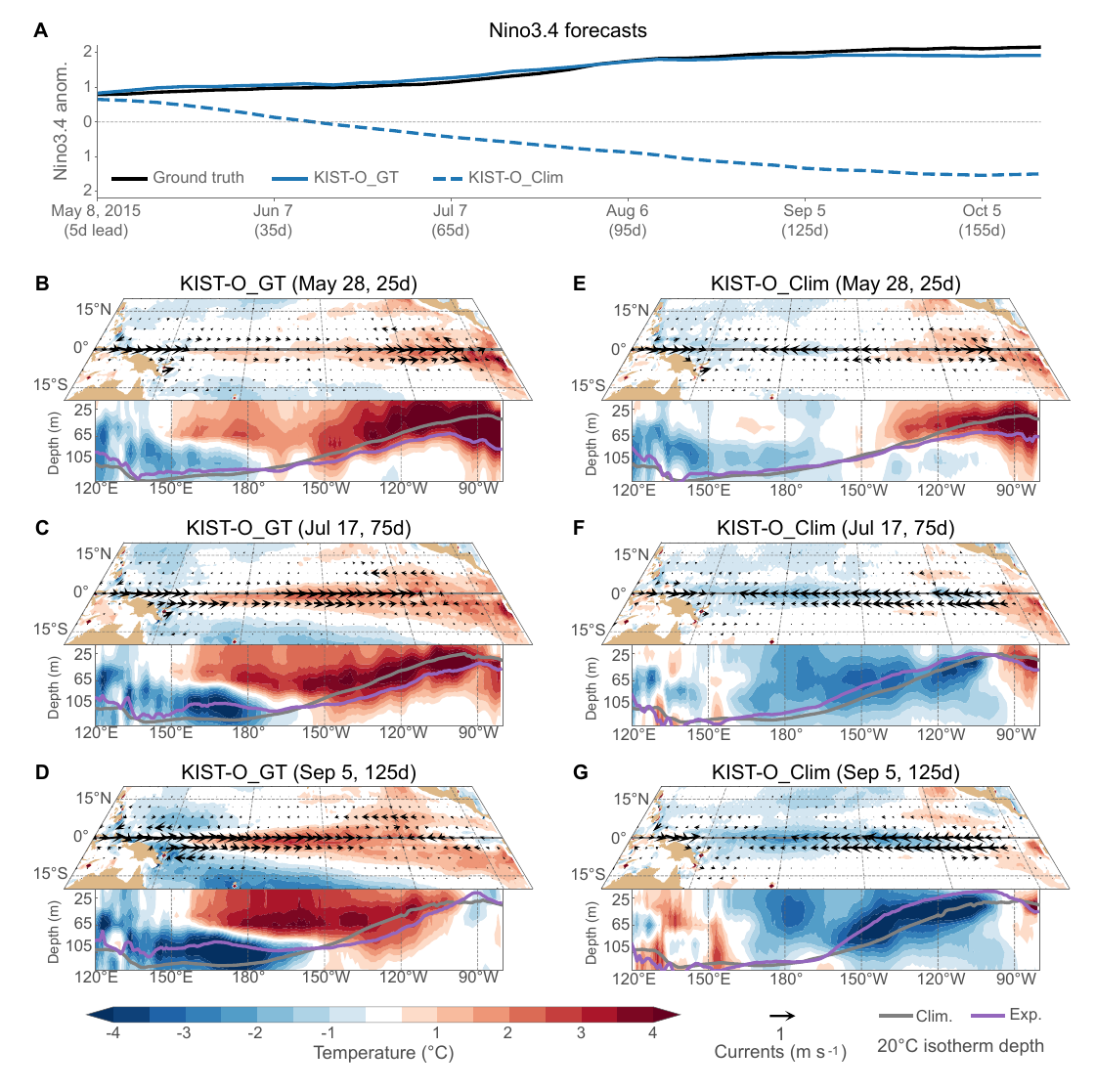}
  \caption{\textbf{2013-2014 Equatorial Pacific temperature evolution under perfect versus climatological surface boundary forcing.} Ocean temperature fields simulated by KIST-Ocean for forecasts initiated from 14 December 2013. Left column (A--C): run with ground truth surface boundary conditions (KIST-O\_GT). Right column (D--F): run with climatological forcing (KIST-O\_Clim). Rows show lead times of 65 days (17 Feb 2014), 95 days (19 Mar 2014), and 125 days (18 Apr 2014). Shading denotes sea surface temperature anomalies (°C, upper strips) and subsurface potential temperature anomalies (°C, lower strips) along the equator. Grey curves mark the climatological 20 °C isotherm depth (T20d); purple curves show the simulated T20d. Black vectors indicate the horizontal wind-stress anomaly applied in KIST-O\_GT. Maps were generated with the Basemap Toolkit (v1.2.0).}
  \label{sifig:si_fig2}
\end{sifigure}

\begin{sifigure}[h!]
  \centering
  \includegraphics[width=0.9\textwidth]{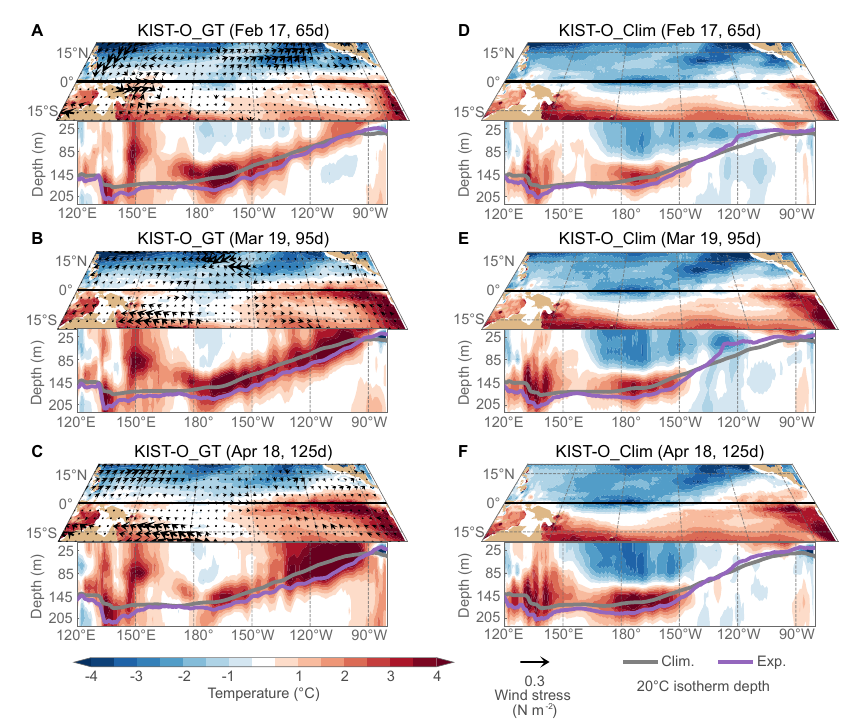}
  \caption{\textbf{Equatorial Pacific thermocline evolution under perfect versus climatological surface boundary forcing.} Longitude-time Hovmöller diagrams of the 20 °C isotherm depth anomaly (m, shading) along 2°S--2°N for forecasts initiated from 14 December 2013. \textbf{A}, KIST-O\_GT: experiment forced with ground truth surface boundary conditions. \textbf{B}, KIST-O\_Clim: experiment forced with climatological surface boundary conditions. Negative (blue) values indicate shoaling of the thermocline; positive (red) values indicate deepening. Black contours (interval 0.4 N m$^{-2}$) show the imposed zonal wind-stress anomaly in KIST-O\_GT. Lead times corresponding to forecast dates are labelled on the y-axis.}
  \label{sifig:si_fig3}
\end{sifigure}

\begin{sifigure}[h!]
  \centering
  \includegraphics[width=0.9\textwidth]{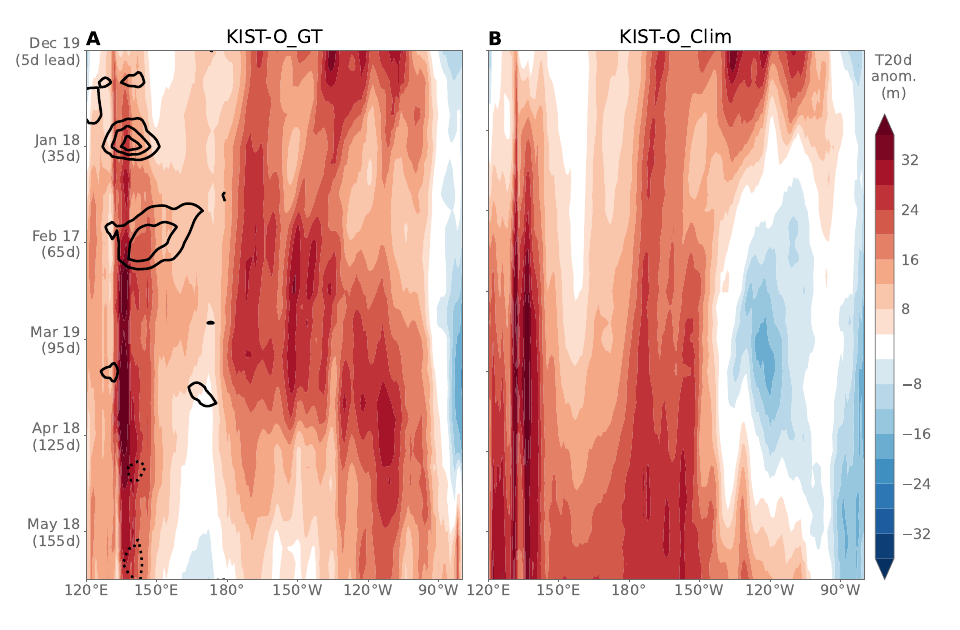}
  \caption{\textbf{2015 Equatorial Pacific temperature evolution under perfect versus climatological surface boundary forcing.} Forecasts are initiated from the ocean state on 3 May 2015. Two surface forcing configurations are compared: KIST-O\_GT (ground truth forcing) and KIST-O\_Clim (climatological forcing). \textbf{A}, Time series of the Nino3.4 (defined by averaging the SST over the region 170°--120°W and 5°S--5°N) anomaly (°C) from 8 May to 19 November 2015 (lead times 5--155 days). Black: GODAS (ground truth); blue solid: KIST-O\_GT; blue dashed: KIST-O\_Clim. \textbf{B--D}, KIST-O\_GT forecasts at lead times of 25 days (28 May 2015), 75 days (17 Jul 2015), and 125 days (5 Sep 2015). \textbf{E--G} Corresponding KIST-O\_Clim forecasts. For each panel, shading in the upper strip shows the sea surface temperature anomaly (°C), and shading below presents the subsurface potential temperature anomaly along the equator. Black vectors denote horizontal ocean current anomalies (m s$^{-1}$). Maps were generated with the Basemap Toolkit (v1.2.0).}
  \label{sifig:si_fig4}
\end{sifigure}

\begin{sifigure}[h!]
  \centering
  \includegraphics[width=0.9\textwidth]{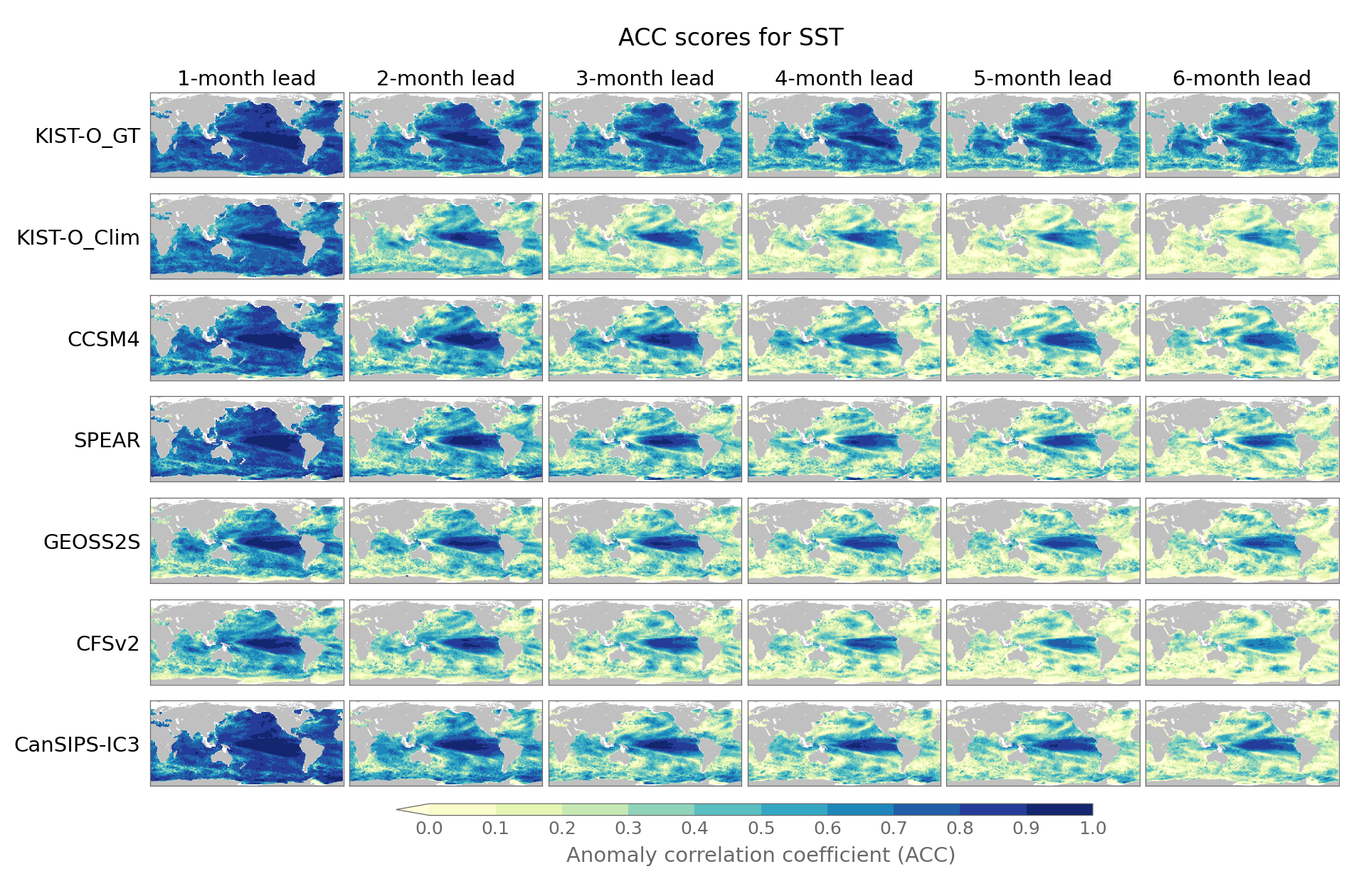}
  \caption{\textbf{Comparison of the horizontal distributions of the ACC for monthly SST predictions by KIST-Ocean and NMME over the period 2015--2022.} Each row (from top to bottom) represents the ACC of KIST-O\_GT, KIST-O\_Clim, COLA-RSMAS-CCSM4, GFDL-SPEAR, NASA-GEOSS2S, NCEP-CFSv2, and CanSIPS-IC3, respectively. From left to right, each column represents predictions from a one-month to six-month lead time. Maps were generated using the Basemap Toolkit (v1.2.0).}
  \label{sifig:si_fig5}
\end{sifigure}

\begin{sifigure}[h!]
  \centering
  \includegraphics[width=0.9\textwidth]{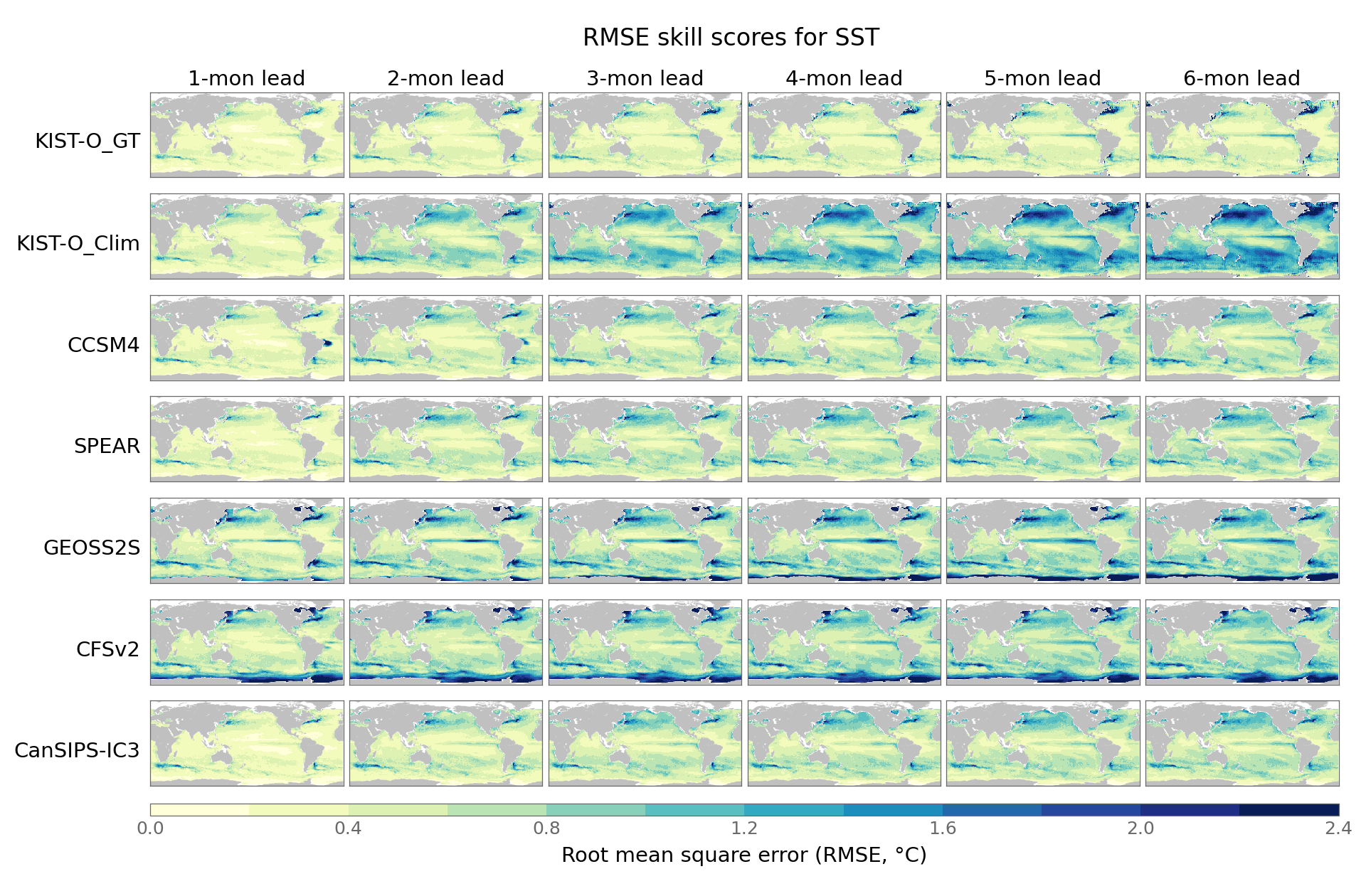}
  \caption{\textbf{Comparison of the horizontal distribution of the root mean square error (RMSE) for monthly SST predictions by KIST-Ocean and NMME models over the period 2015--2022.} Each row (from top to bottom) represents the RMSE (unit: °C) of KIST-O\_GT, KIST-O\_Clim, COLA-RSMAS-CCSM4, GFDL-SPEAR, NASA-GEOSS2S, NCEP-CFSv2, and CanSIPS-IC3, respectively. From left to right, each column represents predictions from a 1-month lead to 6-month lead. Maps were generated with the Basemap Toolkit (v1.2.0).}
  \label{sifig:si_fig6}
\end{sifigure}

\begin{sifigure}[h!]
  \centering
  \includegraphics[width=0.9\textwidth]{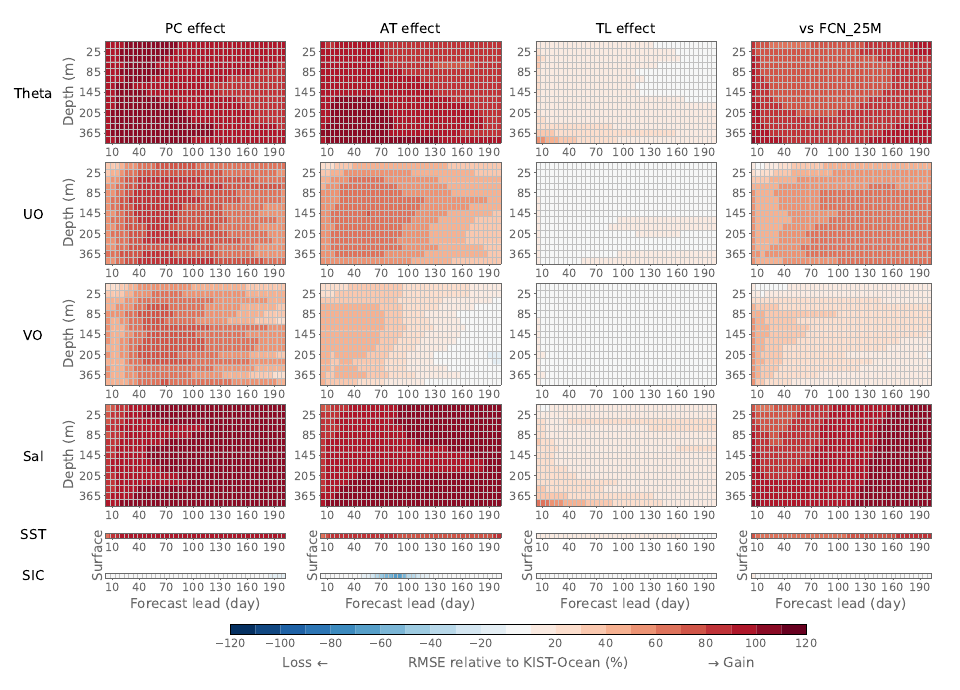}
  \caption{\textbf{Scorecard of the differences in RMSE skills between the original version of the KIST-Ocean model and versions with key algorithms removed, computed over the period 2014--2023.} From left to right, each column corresponds to the model without partial convolution (PC effect), without adversarial training (AT effect), without transfer learning (TL effect), and the model based on FourCastNet (FCN\_25M), respectively. From top to bottom, each row shows the normalized RMSE for potential temperature (Theta), zonal current (UO), meridional current (VO), salinity (Sal), sea surface temperature (SST), and sea ice concentration (SIC), respectively.}
  \label{sifig:si_fig7}
\end{sifigure}

\clearpage
\backmatter

\bibliography{references}

\end{document}